\documentclass[reprint,aps,prfluids,showkeys,twocolumn,superscriptaddress,longbibliography]{revtex4-1}

\usepackage{graphicx}     
\usepackage{subfigure}
\usepackage{hyperref}
\usepackage{bm}
\usepackage{ae}
\usepackage{aecompl}
\usepackage{amsmath}
\usepackage{amssymb}
\usepackage{epsfig}
\usepackage{psfrag}
\usepackage{subfigure}
\usepackage{textcomp}

\newcommand{\Eq}[1]{Eq.~(\ref{eq:#1})}
\newcommand{\rpar}{{\bf x}} 

\newcommand{\hh}{h}                
\newcommand{\z}{z}                
\newcommand{\zsq}{\z_{\rm i-f}}    
\newcommand{\zqv}{\z_{\rm f-v}}    
\newcommand{\SqL}{i-f}    
\newcommand{\qLV}{f-v}    
\newcommand{\lv}{_{lv}}    
\newcommand{\sv}{_{sv}}    

\usepackage{color}

\begin{document}

\title{Rounded Layering Transitions on the Surface of Ice}





\author{Pablo Llombart$^{\dag,\ddag}$, Eva G. Noya$^\ddag$, David N.
Sibley$^\S$, Andrew J. Archer$^\S$ and Luis G. MacDowell$^\dag$*}
\affiliation{$^\dag$Departamento de Qu\'{\i}mica-F\'{\i}sica (Unidad de I+D+i
   Asociada al CSIC), Facultad de Ciencias Qu\'{\i}micas, Universidad Complutense de Madrid, 28040 Madrid, Spain, \\  $^\ddag$Instituto de Qu\'{\i}mica F\'{\i}sica Rocasolano
, CSIC, Calle Serrano 119, 28006 Madrid, Spain and \\
  $^\S$Department of Mathematical Sciences, Loughborough University,
  Loughborough LE11 3TU, United Kingdom.
}

\email[]{lgmac@quim.ucm.es}




\begin{abstract}
   Understanding the wetting properties of premelting films requires
   knowledge of the film's equation of state, which is not usually available.
   Here we calculate the disjoining pressure curve of premelting films,
   and perform a detailed thermodynamic characterization of premelting
   behavior on ice.  Analysis of the density profiles reveals the signature 
   of weak layering phenomena, from one to two and from two to three water
   molecular layers. However, disjoining pressure curves,
   which closely follow expectations from a renormalized mean field
   liquid state theory, show that there are no layering phase transitions
   in the thermodynamic sense along the sublimation line. Instead, 
   we find that transitions at mean field level are rounded
   due to capillary wave fluctuations.  We see signatures that 
   true first order layering transitions could arise at low temperatures, 
   for pressures between the metastable line of water/vapor coexistence and 
   the sublimation line. The extrapolation of the disjoining pressure curve
   above water vapor saturation displays a true first order phase transition
from a thin to a thick film consistent with experimental observations.
\end{abstract}



\maketitle


Understanding the properties of the surface of ice is of crucial importance
in many important phenomena, such as the growth of snowflakes \cite{libbrecht17},
the freezing and melting rates of ice at the poles \cite{dash06}, or the
scavenging of trace gases on ice particles \cite{abbat03}. Interestingly,
close to the triple point the ice surface is known to exhibit surface
premelting, i.e., the appearance of  a loosely defined quasi-liquid layer \cite{furukawa87}, 
which is expected to have an important effect on crystal growth rates \cite{kuroda82},
adsorption \cite{abbat03}, friction \cite{weber18}, and many 
other properties \cite{slater19,nagata19}.

\begin{figure*}[htb!]

   \subfigure[]{\label{perfil_liq1_basal}
   \includegraphics[width=0.29\textwidth,height=0.35\paperwidth,keepaspectratio]{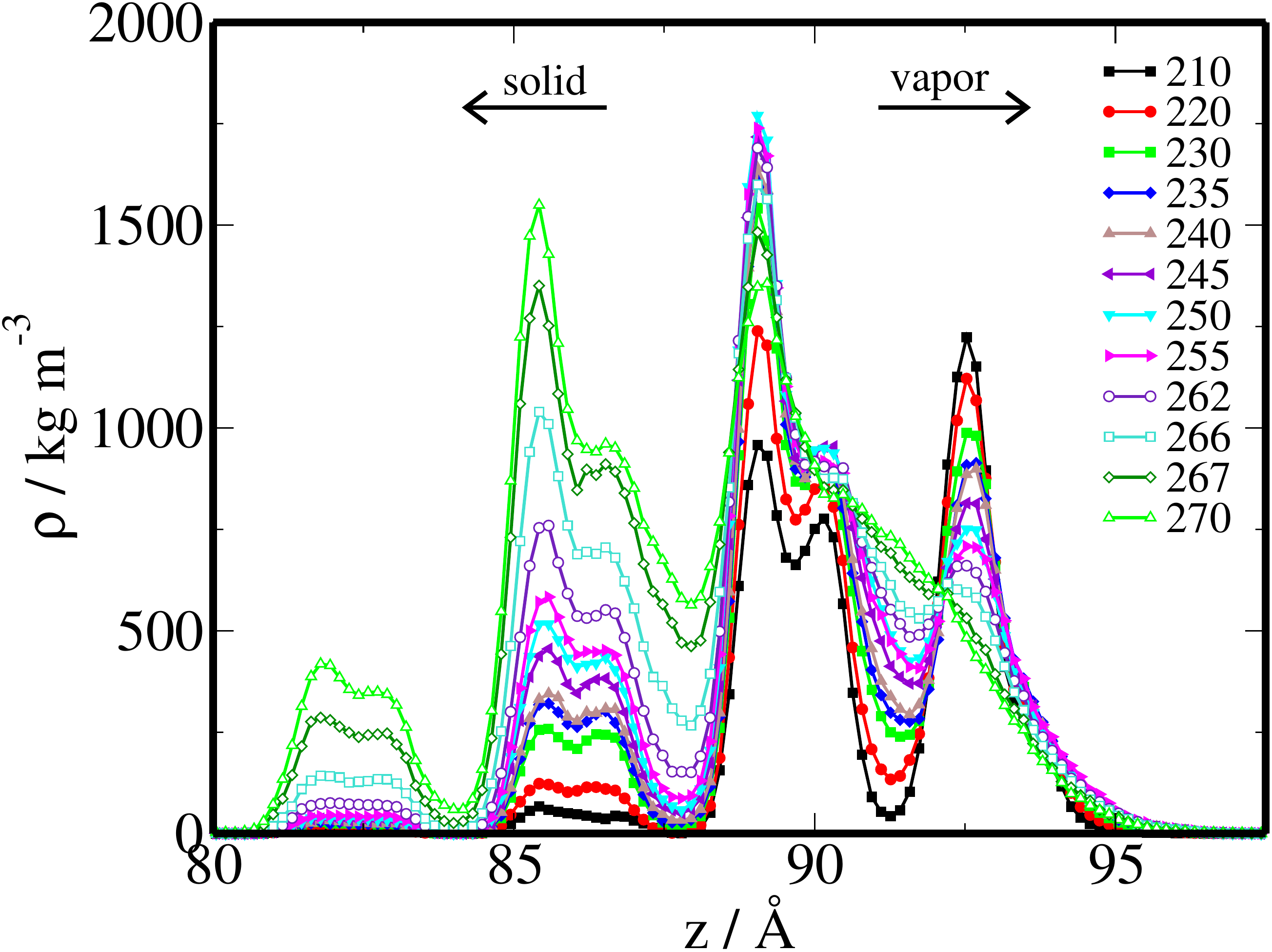}}
   \subfigure[]{\label{perfil_liq2_basal}
   \includegraphics[width=0.29\textwidth,height=0.35\paperwidth,keepaspectratio]{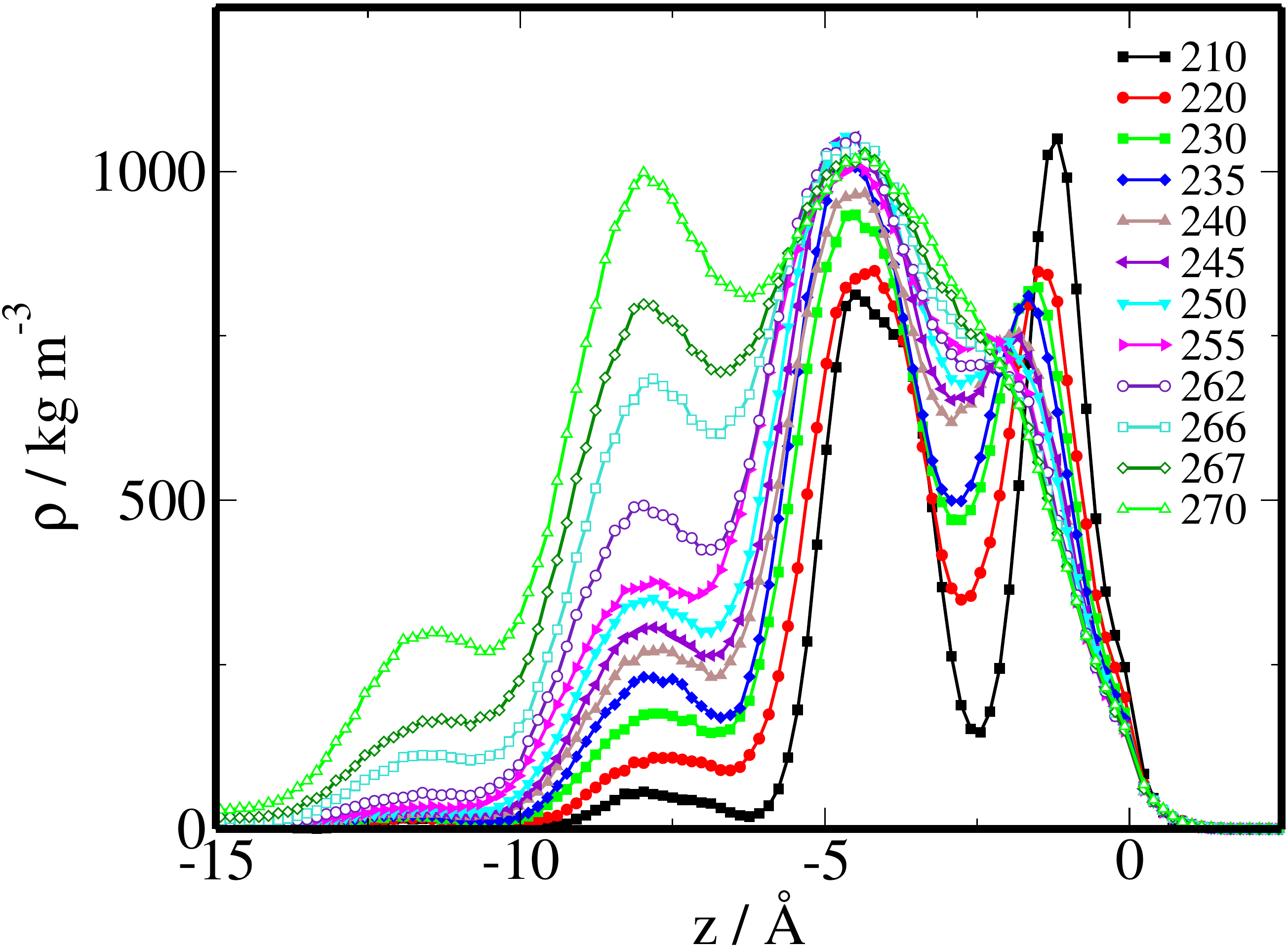}}
   \subfigure[]{\label{perfil_sol2_basal}
	  \includegraphics[width=0.3\textwidth,height=0.35\paperwidth,keepaspectratio]{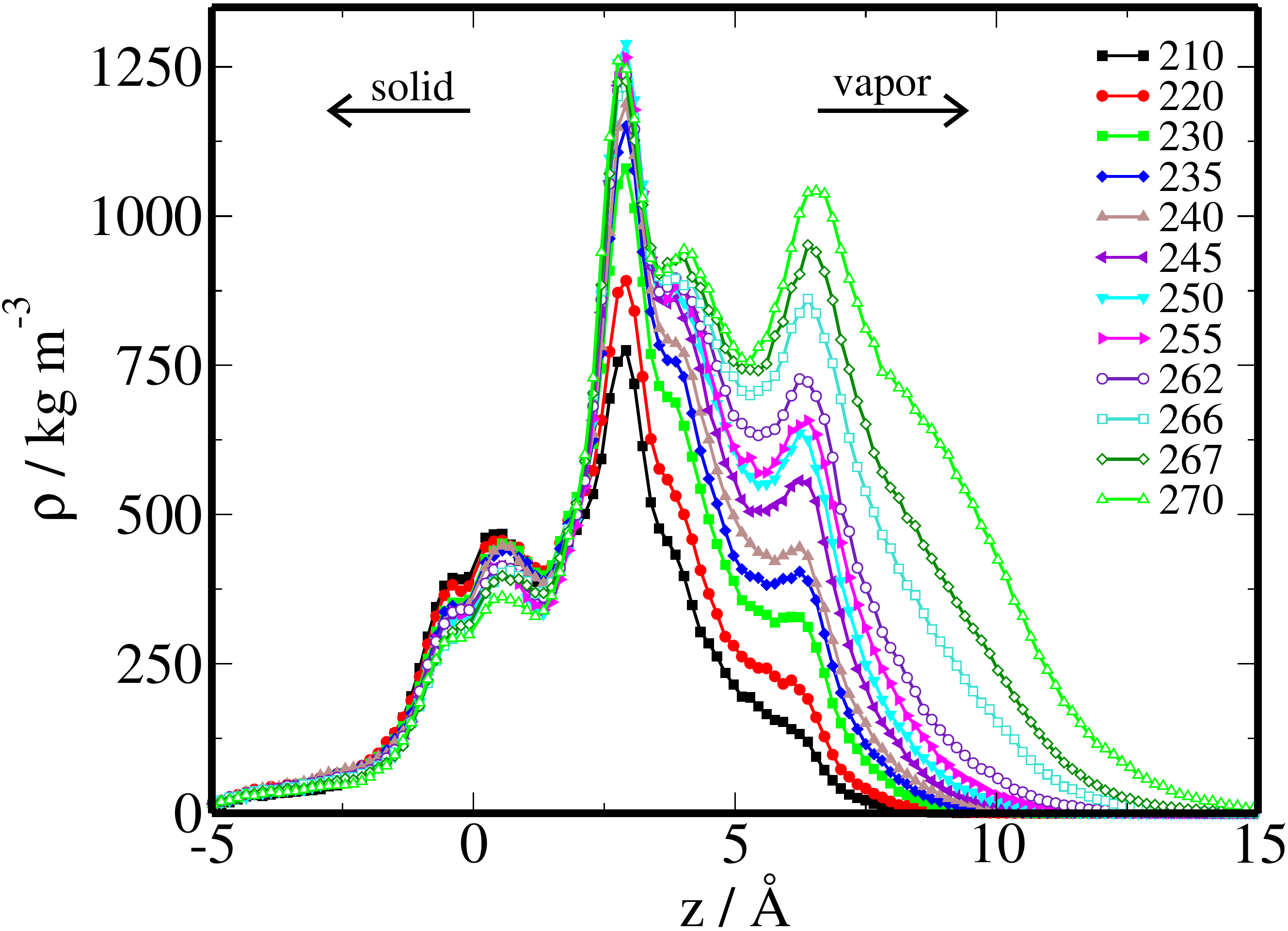}}
	  \vspace*{-0.2cm}
   \caption{Density profiles of liquid-like molecules for the
basal interface as measured relative to (a) the laboratory reference frame
   (b)   the \qLV~surface and 
 (c)  the \SqL~surface. 
    Since the liquid-like layer has finite thickness, the
    profiles vanish  at the solid phase to the left and the vapor phase to the right
    of the z axis in these figures.
 %
}
	 \label{perfiles}
 \end{figure*}
But despite its significance, and  great experimental 
progress \cite{furukawa87,dosch96,wei01,bluhm02,sadtchenko02,murata16,smit17,sanchez17},
the characterization  of ice premelting has remained
a longstanding matter of debate \cite{li07b,michaelides17,slater19}.
As the controversy regarding the film thickness could
start to clarify with agreement between widely different experimental
techniques \cite{bluhm02,sadtchenko02,conde08,constantin18,mitsui19}, new
and exciting observations have been made regarding the properties
of the premelting film, both at \cite{sanchez17,pickering18,qiu18}, 
and off ice/vapor 
coexistence \cite{murata16}. Sanchez et al.\ performed
a Sum Frequency Generation experiment (SFG) of the ice surface along the
sublimation line and found 
evidence of a discrete bilayer melting transition \cite{sanchez17,michaelides17}. 
Off coexistence, experiments \cite{murata16} have found a discontinuous 
transition from a thin to a thick premelting film occurring at 
water-vapor supersaturation that is often known as  frustrated complete
wetting \cite{shahidzadeh98,indekeu99,mueller01b,macdowell05}.
An appealing interpretation is to view both phenomena 
as manifestations of layering effects and renormalization similar to
those studied in past decades for simple model 
systems \cite{weeks82, huse84, patrykiejew90, chernov88, ball88, evans92, henderson94, brader01, dijkstra02}.
However, the evidence for layering is not without some controversy
too as other studies also based on SFG suggest that
bilayer melting 
occurs instead in a
continuous fashion \cite{smit17}. 
Interestingly,
computer simulations show that 
the ice surface exhibits patches of premelted ice, whose size increases
continuously as the temperature is increased, also consistent with
a continuous 
build up of the premelting film \cite{hudait17,pickering18,qiu18}.

Here we show that in the range between 230 to 270~K,
the order parameter distributions 
of the main facets of ice
 exhibit the signature of
 two consecutive rounded 
layering transitions. 
This reconciles conflicting experimental
and computer  simulation studies of the equilibrium surface
structure \cite{sanchez17,smit17,hudait17,pickering18,qiu18}.
Extrapolation of our results for the TIP4P/Ice model
according to predictions of liquid state and renormalization theory
indicate that
a genuine first order phase transition
occurs at supersaturation, consistent with experimental observations
off-coexistence \cite{murata16}.
Our results
provide a unified vision of the wetting behavior of premelting
films on ice.

We use the TIP4P/Ice model, which exhibits a melting
temperature of 272~K \cite{abascal05}. To prepare the system
in the solid/vapor coexistence region, we  place
an ice slab of either 1280 or 5120 molecules in vaccuum.
Performing canonical Molecular Dynamics simulations 
\cite{bussi07} with GROMACS,  
the system attains two phase coexistence in a few nanoseconds and
an equilibrated premelting film is formed spontaneously
\cite{conde08,benet16,kling18,llombart19}
(Supplementary Material).
Previously, computer simulation evidence for a layering transition 
of the TIP4P/Ice model has been
discussed in terms of the density profile $\rho(z)$ of the H$_2$O
molecules as a function of the perpendicular distance, $z$ to the 
interface \cite{sanchez17}. 
A more detailed description of layering phenomena in terms of
density profiles is afforded by identifying liquid-like and 
solid-like environments \cite{lechner08,nguyen15}.
Using  the $\bar q_6$ order parameter \cite{lechner08} 
allows us to plot the density
of liquid-like and solid-like molecules, and identify features that
are specific to the premelting layer \cite{benet16,benet19,llombart19}.

The density profile of liquid-like molecules at the 
   $\{0 0 0 1\}$
face (or basal face)
can be
seen in Fig.~\ref{perfiles}(a). At low temperatures, the profile
is highly structured, with the main peaks separated by about 3.5~\AA,
a value very close 
to the preferred lattice spacing of the underlying solid. This, together
with the bilayer structure apparent as a double peak is suggestive of
a rather ordered liquid-like environment. As temperature rises,
the profiles 
transform in a continuous manner up to about 267~K. At this
temperature a qualitative change is observed, 
as the outermost maxima and minima
of the density profile disappear via
an inflection point, leaving the profile with a monotonic decay into
the vapor phase. 

In fact, the strong
stratification of the liquid-like profile is caused to a great extent
by the structure of the underlying ice lattice.
To show this, we describe the premelting liquid layer
in terms of two bounding surfaces separating the quasi-liquid (fluid)
film from bulk ice and bulk vapor \cite{benet16,benet19,llombart19}. For points $\rpar$ along
a flat reference plane, we calculate an ice-fluid (\SqL) surface
as the loci, $\zsq(\rpar)$ of the outermost solid-like molecules. Likewise,
we calculate a fluid-vapor (\qLV) surface, as the loci, $\zqv(\rpar)$
of the outermost liquid-like molecules (see Supplementary Material).
An intrinsic density profile relative to the
local \qLV~interface can then be determined as the density of atoms
at a distance $z-\zqv(\rpar)$. The picture that emerges
[Fig.~\ref{perfiles}(b)] shows a much less layered structure, confirming that the strong
stratification of $\rho(z)$ shown in Fig.~\ref{perfiles}(a) is largely the result
of structural correlations conveyed by the solid phase. The profile
evolves continuously up to 264~K, with the outermost minimum and maximum
again disappearing across
an inflection point at about $z-\zqv(\rpar)=-2.2$~\AA,
as the decay of the density profile towards the vapor phase
takes the monotonic form characteristic of a liquid-vapor interface.

The same insight can be found by plotting the intrinsic density
profile of liquid-like molecules as measured relative to the
fluctuating \SqL~surface [Fig.~\ref{perfiles}(c)],
which is again 
 less
stratified than the absolute density profile.
This 
profile also evolves in a rather continuous fashion,
but, surprisingly, it does not show any particular signature of layering
between 265 and 270~K as observed previously.
Rather,
in this case one notices the appearance of a maximum
at about 6.7~\AA\ which occurs via an inflection point
between 230 and 235~K. 

Whilst the details of the liquid-like density profiles at the prism facet
   $\{1 0 \overline{1} 0 \}$
are very different, the
main features observed at the basal facet are also found here, with 
intrinsic density profiles that are again very much rounded relative to the
strongly stratified absolute density profile.
A close inspection 
shows inflection points appear between 265 and 270~K 
and between 240 and 250~K
(Supplementary Material).

In order to clarify the process of bilayer melting further,
we need to resort to a more illuminating order parameter than
the density profiles of liquid-like molecules. Based on the intrinsic
\SqL~and \qLV~surfaces, we can define an
instantaneous local film
thickness  as $\hat{\hh}(\rpar) = \zqv(\rpar)-\zsq(\rpar)$.
We exploit this local parameter
to calculate the mean film thickness $\hh$ 
after lateral and canonical averaging over the simulation run.

\begin{figure*}[htb!]
\centering

 \subfigure[]{\label{fig:espesor_big_basal}
\centering
      \includegraphics[width=0.25\paperwidth,keepaspectratio]{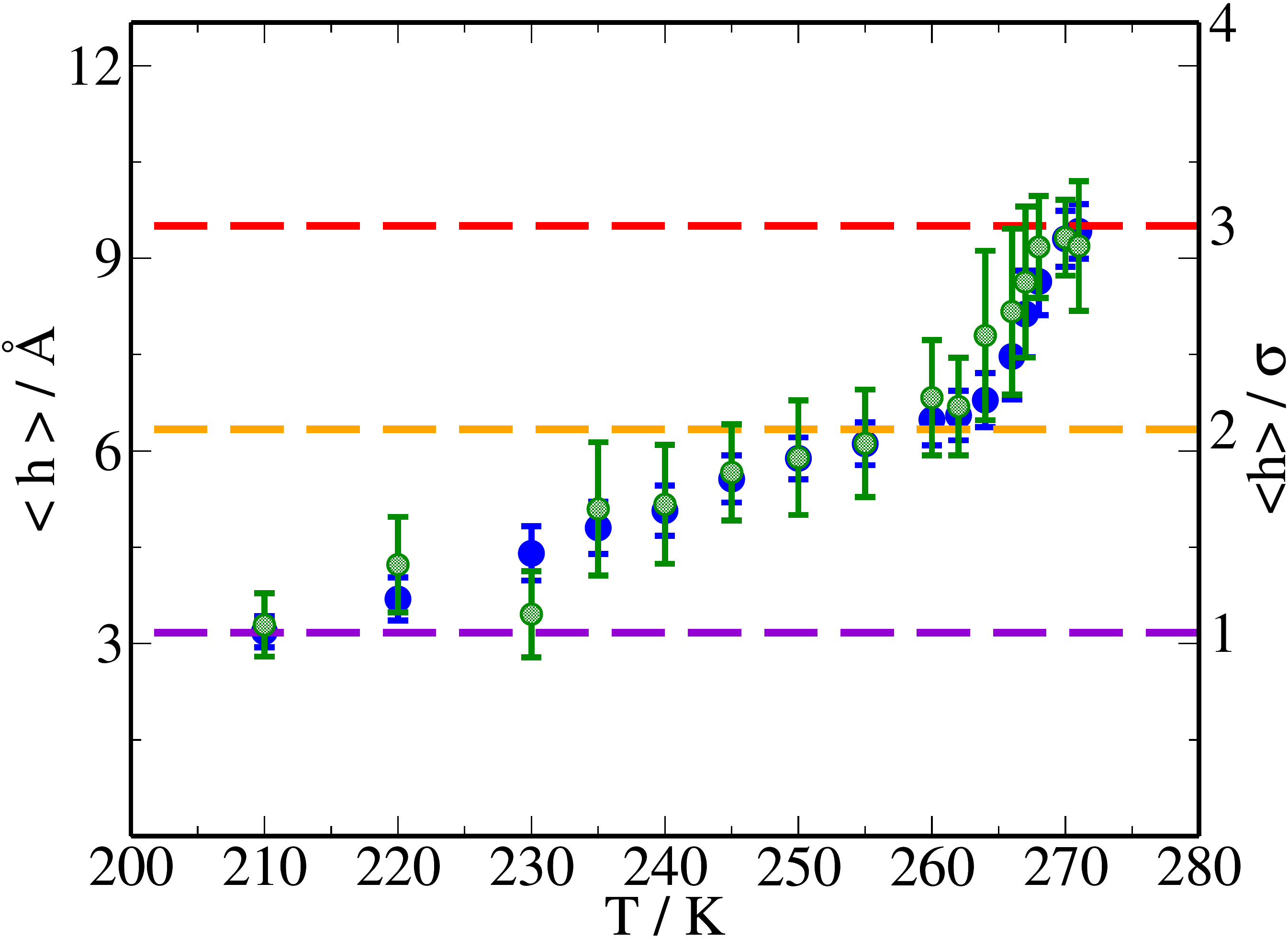}}
 \subfigure[]{\label{fig:disjoining_basal}
\centering
      \includegraphics[width=0.25\paperwidth,height=0.18\paperwidth]{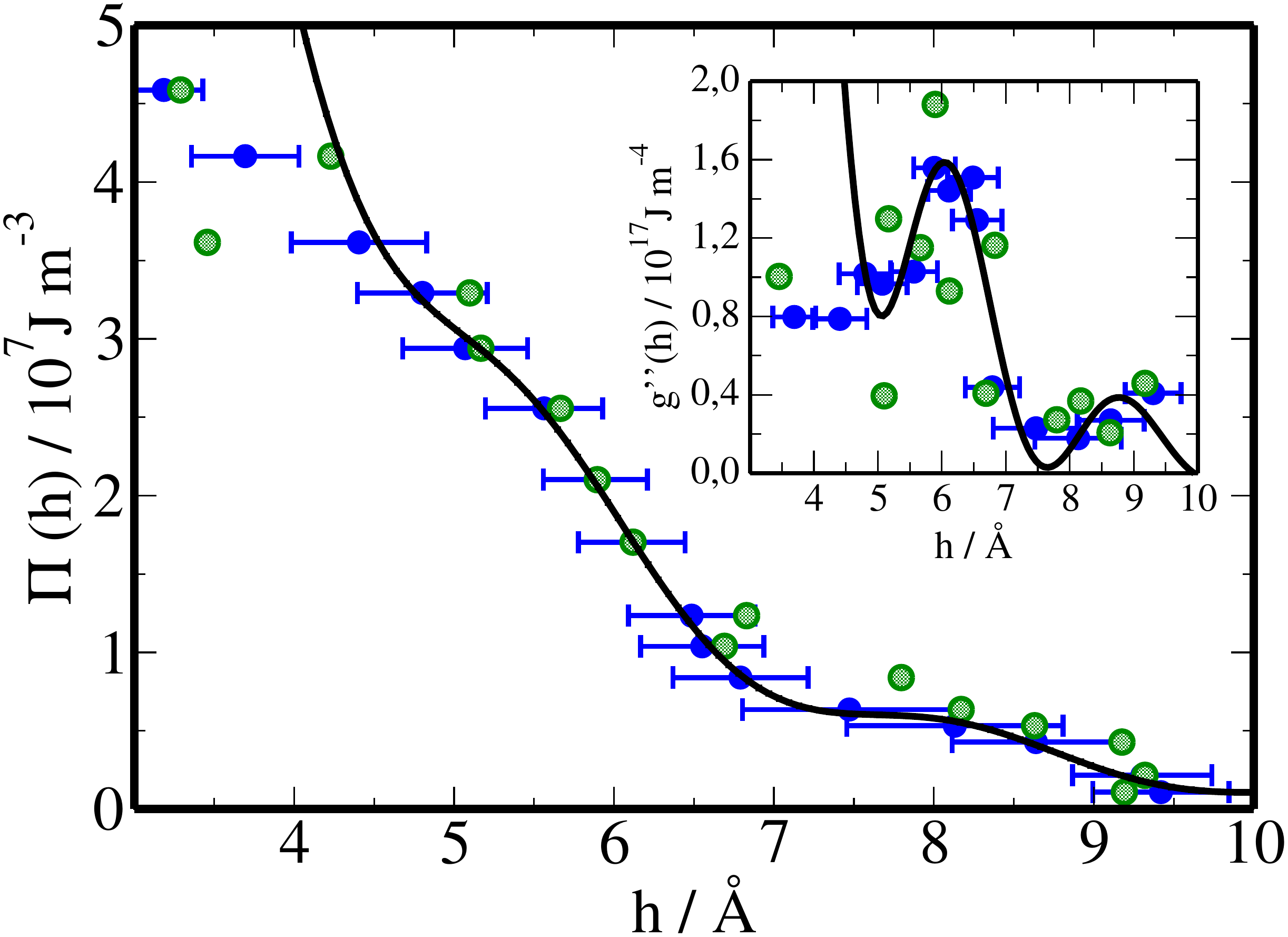}}
\subfigure[]{\label{fig:g_de_h_b}
\centering
      \includegraphics[width=0.23\paperwidth,height=0.18\paperwidth]{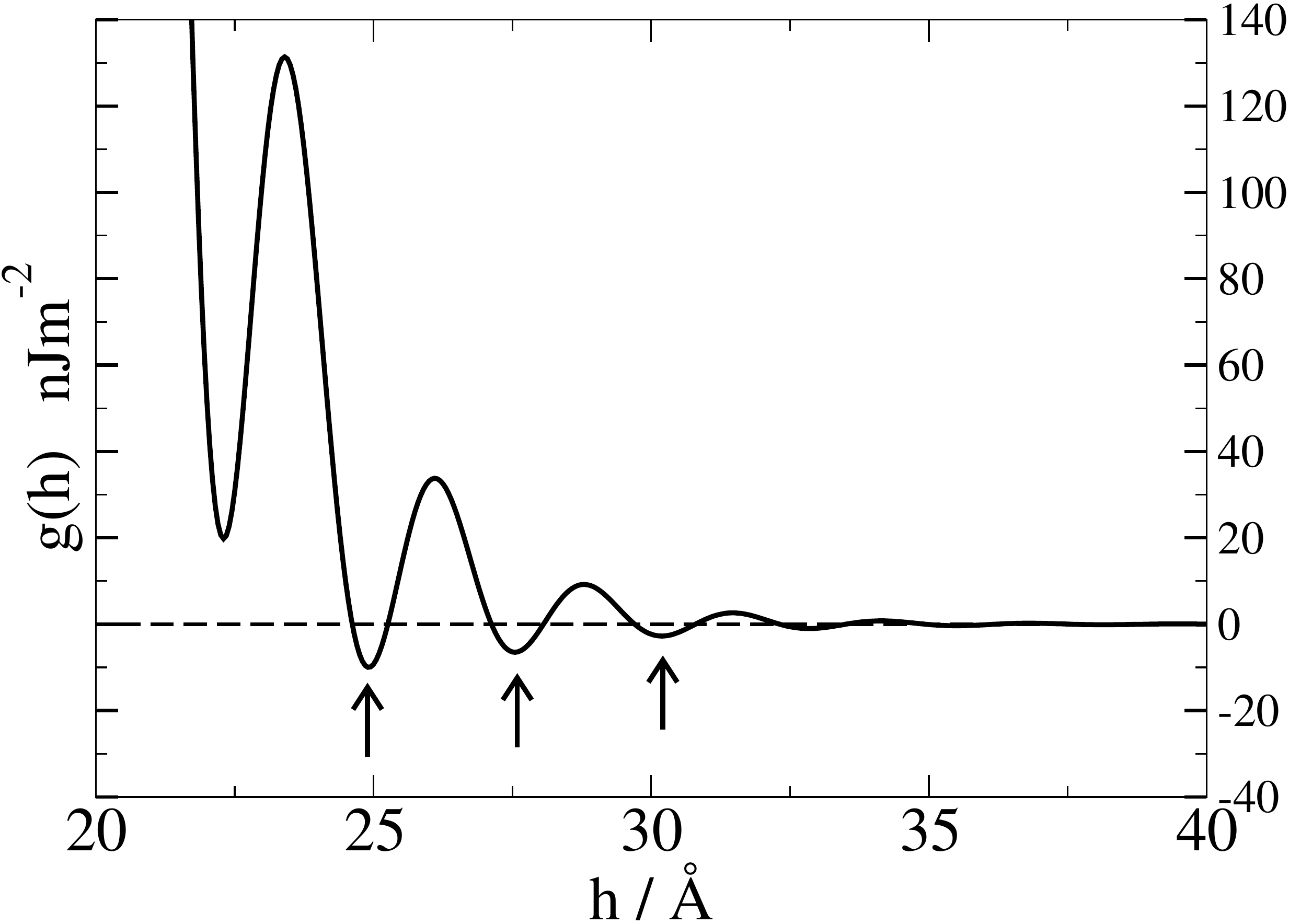}}
	  \vspace*{-0.2cm}
	\caption{Film thickness (a), disjoining pressure (b), and interface
potential (c) for the basal facet.
      Results are shown for two system sizes with 
      $n_{x}\times n_{y} =5120$ (blue filled circles) and  
      $n_{x}\times n_{y} =1280$ (green hollow circles) molecules.  
	(a) The dashed lines indicate discrete film heights in units 
	of the molecular diameter.  (b) 
        The lines are a fit of \Eq{grn} to the blue symbols
	  for $\hh>4.0$~\AA. The insets show inverse surface
	  susceptibilities as obtained from numerical deriviatives (symbols) and
	  from analytical derivatives of the fits (lines). Error bars
	  are shown here only for the system fitted to \Eq{grn}.
       Panel (c) shows the decay of the interface potential at 
intermediate distances as extrapolated from the fit to \Eq{grn}. 
Arrows indicate states that can transition
at supersaturation, i.e.\ to frustrated complete wetting states.
   }
	\label{espesores}
 \end{figure*}

Fig.~\ref{fig:espesor_big_basal} 
depicts the results obtained for the basal 
plane.
The film thickness grows from about one molecular layer at $T$=210~K,
to about three molecular layers at $T$=270~K in a continuous fashion,
with no clear evidence of a first order layering transition. 
Results obtained recently for the monoatomic water 
model (mW) show a similar trend
but appear smoother as they are sampled in the grand canonical
ensemble, which allows for larger film thickness fluctuations 
\cite{pickering18,qiu18}.
We provide a full thermodynamic characterization of wetting properties
by calculating the disjoining pressure of the film, $\Pi(\hh)=-d g(\hh)/d\hh$,
where $g(\hh)$ is the binding potential, which accounts
for the pressure difference between the adsorbed liquid film and
the bulk vapor of equal chemical potential,
i.e.~\cite{derjaguin87,henderson05,benet14b}:
\begin{equation}\label{eq:derjaguin}
   \Pi(\hh) = p_v(\mu,T) - p_l(\mu,T).
\end{equation} 
In practice, $\Pi(\hh)$ is to an adsorbed liquid film
as the Laplace pressure is to a liquid droplet. In particular,
the vapor pressure of an ideal gas in equilibrium with a film of 
thickness  $\hh$ is given in a manner analogous to the Kelvin-Laplace equation as
$p_v = p_{v,w} e^{-\beta \Pi(\hh)/\rho_w}$, where $p_{v,w}$ is the saturated vapor pressure over water,
$\rho_w$ is the bulk liquid density at coexistence and $\beta=(k_B T)^{-1}$.  An equilibrium film thickness for  water adsorbed at
the ice/vapor interface can be meaningfully defined only along the sublimation line, 
where the vapor pressure equals the saturated vapor pressure over ice, $p_{v,i}$. Accordingly,
the above equation becomes $p_{v,i} = p_{v,w} e^{-\beta \Pi(\hh)/\rho_w}$. It follows that
using accurate coexistence vapor pressures, and
the corresponding equilibrium film thickness $\hh(T)$ along
the sublimation line, we can readily determine $\Pi(\hh)$ (Supplementary Material).
The significance of this result can be hardly overemphasized.
By exploiting the data $\hh(T)$ 
at solid/vapor coexistence, we can now determine the film height
of the premelting film at arbitrary temperature {\em and pressure},
by merely solving \Eq{derjaguin} for $\hh$ \cite{sibley19}. This
is a required input in theories of
premelting \cite{kuroda82,nenow86,wettlaufer19}.

Results for 
the basal  surface are shown in
Fig.~\ref{fig:disjoining_basal}.
The disjoining pressure curve measured up to $\hh=10$~\AA~
exhibits a monotonic behavior, but with a clear
damped oscillatory decay at positive disjoining pressure.
In contrast, a system exhibiting first order layering transitions
exhibits sinusoidal oscillations in mean field, or alternatively,
an equal areas Maxwell construction with a segment of zero slope
beyond mean field. We confirm the presence of the oscillatory behavior
from plots of the derivative (obtained numerically) -- see inset.
Maxima of the inverse susceptibility, 
$\chi_{\parallel}^{-1}=d^2g/d\hh^2$,
which characterizes parallel correlations,
indicate enhanced stability at preferred film thicknesses of 
$\hh=6$~\AA~for the basal plane and at $\hh=5.4$~\AA~for the
prism plane (Supplementary Material), 
but this is not a sufficient criteria for
a thermodynamic phase transition.
Thus, it appears that at a mean field level there could be
layering which is washed away upon renormalization to larger
length scales, as suggested from the study of capillary wave
fluctuations \cite{chernov88,henderson94}.

Liquid state theory provides an expansion for the renormalized
interface potential $g(\hh)$ of a premelting film dominated by short
range structural forces. 
To leading order in $\hh$, this gives:
\begin{equation}\label{eq:grn}
g_R(\hh) = A_2 e^{-\kappa\hh} - A_1 e^{-\kappa_R\hh} \cos(k_{z,R}\hh),
\end{equation} 
where the amplitudes $A_1$ and $A_2$ depend on $T$, while
$\kappa$ and $\kappa_R$ are inverse length scales
that characterize the decay of the pair correlations in the liquid.
Their values are renormalized from those one would expect
on the basis of mean field theory, so that e.g.\ $k_{z,R}<k_z$, where
$k_z$ is the wavenumber corresponding to
the maximum of the bulk liquid structure
factor \cite{chernov88, evans92, henderson94, evans94, hughes17}.

This result is an asymptotic form, and is not expected to
hold for small thicknesses of barely one molecular diameter. 
However,
a fit 
of $\Pi(\hh) = -d g_R(\hh)/d\hh$ as predicted by \Eq{grn}
for all $\hh>4.0$~\AA~
exhibits an excellent agreement with simulation data.
Moreover, from the fitting parameters we can determine
the wetting behavior at short and intermediate $\hh$.
For the basal facet ($\kappa=0.61$~\AA$^{-1}$, $\kappa_R=0.43$~\AA$^{-1}$ and
$k_{z,R}=2.36$~\AA$^{-1}$) we find $\kappa_R<\kappa$, so that
$g(\hh)$ exhibits an absolute minima at intermediate distances. 
This implies incomplete wetting  of the premelting film,
as expected for facets below the roughening
temperature of the solid/melt 
interface \cite{chernov88,evans92,henderson94}.
For the
prism plane ($\kappa=0.62$~\AA$^{-1}$, $\kappa_R=0.90$~\AA$^{-1}$ and
$k_{z,R}=3.14$~\AA$^{-1}$)
we find in contrast that $\kappa_R>\kappa$, so that
$g(\hh)$ exhibits a monotonous decay. This leads to
complete wetting of the premelting film 
as a result of a fluctuation dominated wetting transition \cite{chernov88}.
Of course, for very large $\hh$ the decay rate of $\kappa_R$ is
not expected to depend on the crystal facet.
However, for  intermediate values of $\hh$ 
of interest here the decay  need not be exactly the same, 
since the renormalized quantities
depend on details of the interface potential at
short range. 
In practice, at sufficiently large distance van der Waals forces
with algebraic decay will favor incomplete wetting 
irrespective of the surface plane involved \cite{elbaum91b}.

\begin{figure*}[htb!]

  \subfigure[]{\label{p_global_b}
     \includegraphics[width=0.29\textwidth,height=0.35\textwidth,keepaspectratio]{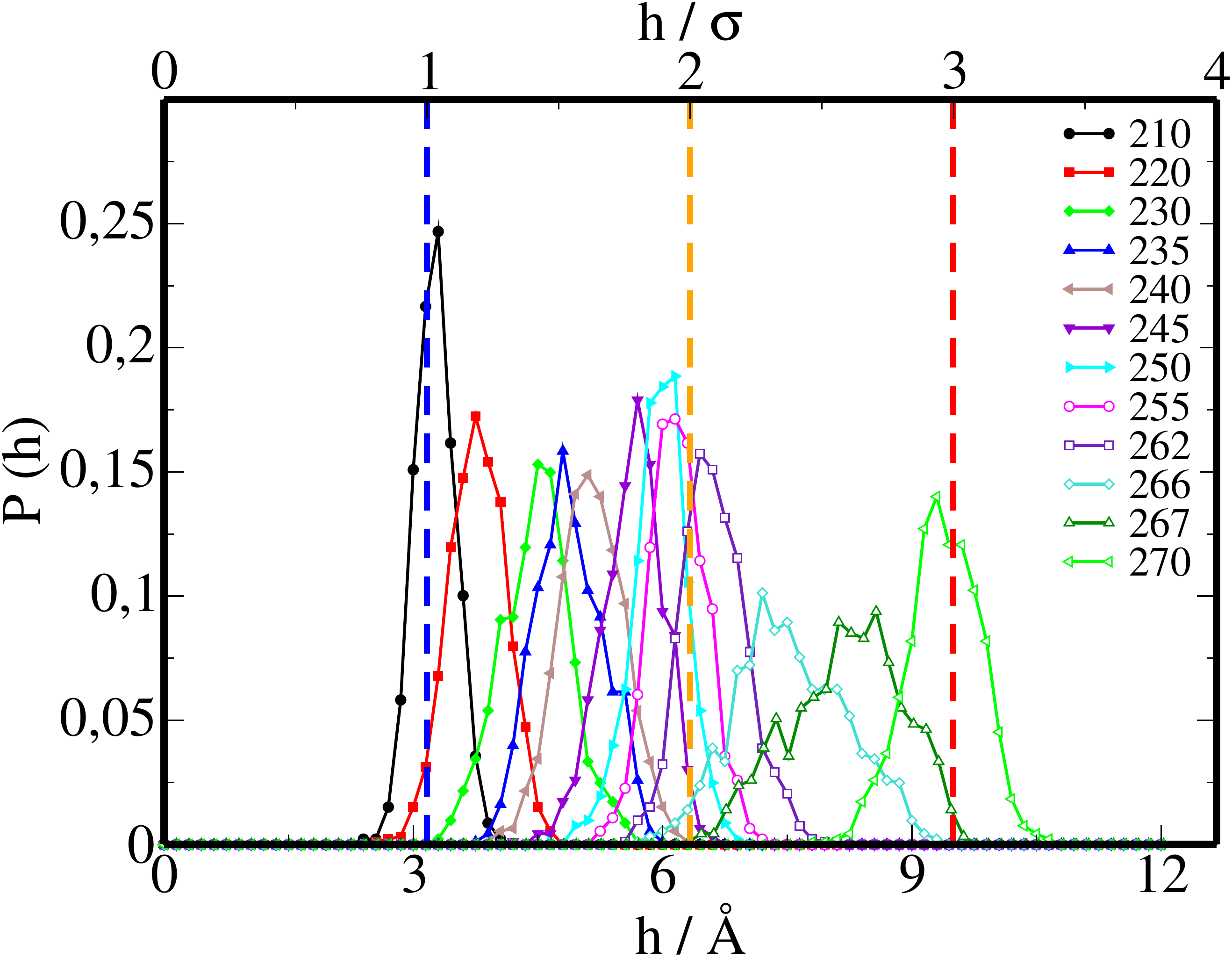}}
  \subfigure[]{\label{p_semilocal_b}
     \includegraphics[width=0.29\textwidth,height=0.35\textwidth,keepaspectratio]{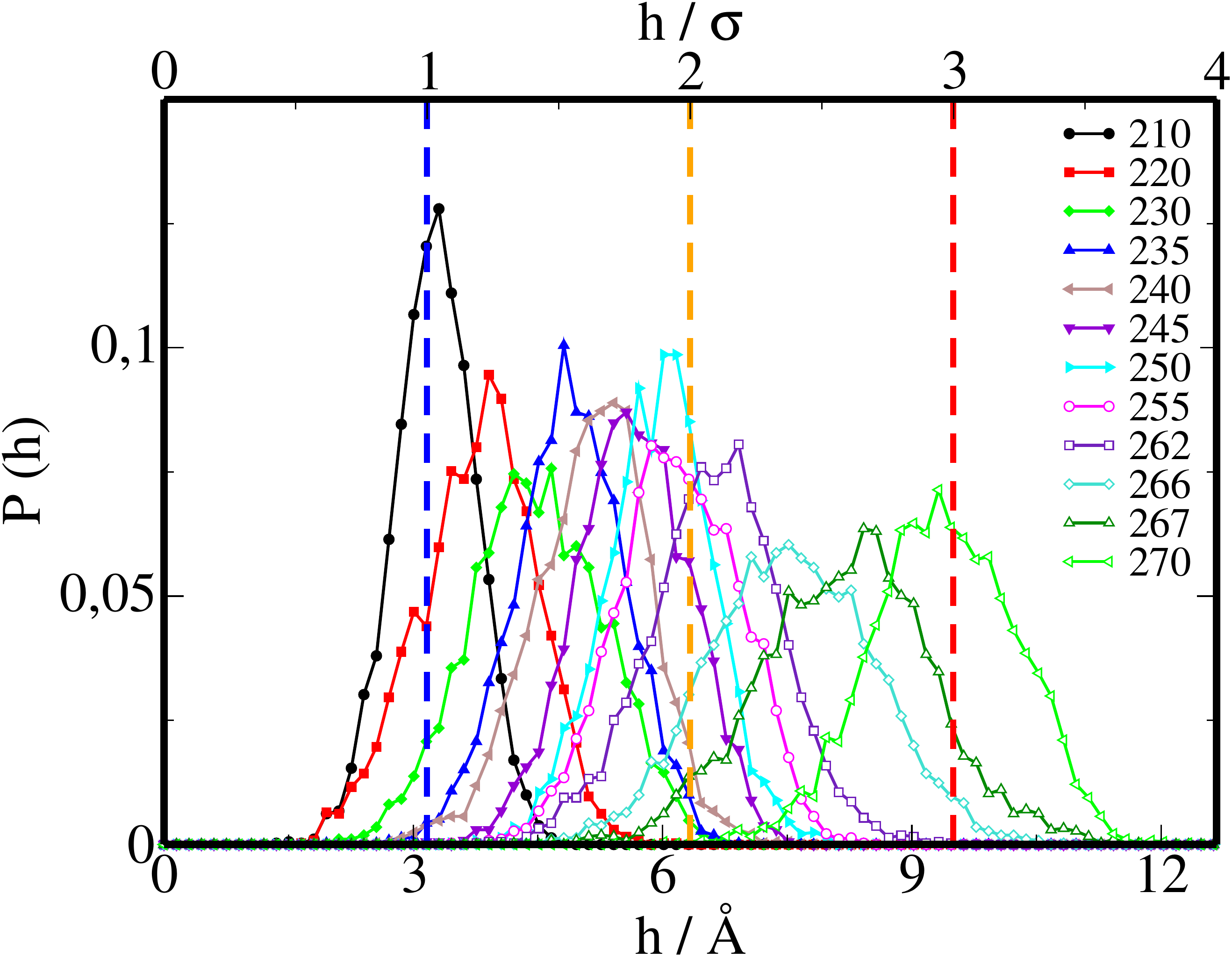}}
  \subfigure[]{\label{p_local_b}
     \includegraphics[width=0.29\textwidth,height=0.35\textwidth,keepaspectratio]{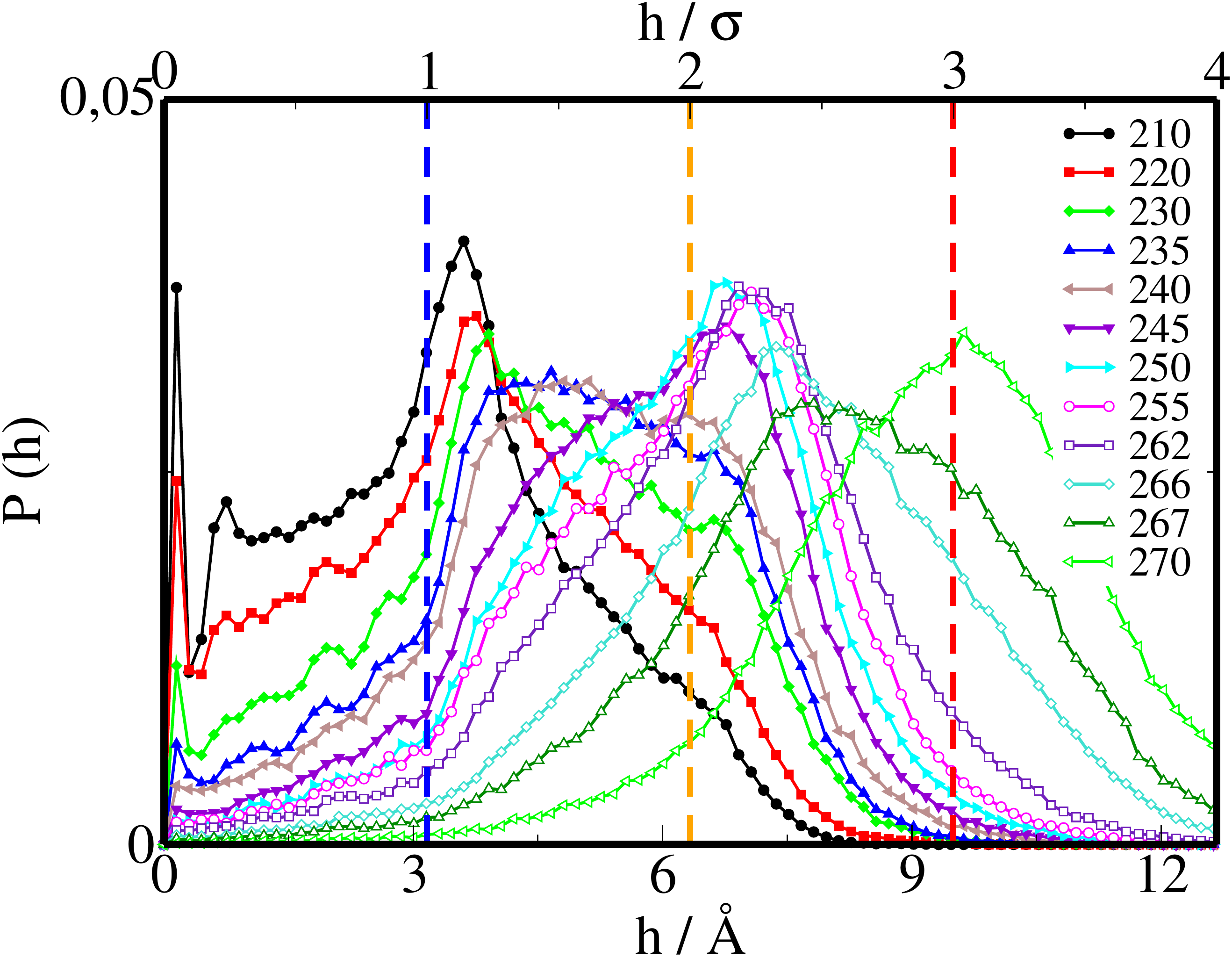}}
	  \vspace*{-0.2cm}
	\caption{System size analysis of thickness distributions at the
	basal plane.  (a) probability distribution of the
global film thickness, $h$. (b) probability distribution of the partial
film thickness $h_{1/4}$. (c)  probability distribution of the local order
parameter $h(\rpar)$. Dashed vertical lines show the film thickness in units of
the molecular diameter. Results are shown for temperatures in the range from
$T$=210 K to $T$=271 K, with color code as indicated on the figure. 
 }
 \label{fig:dist_basal}
\end{figure*}

To clarify whether the layering  is consistent with either
a continuous or  a first order phase transition, we perform a
block analysis of the film thickness distributions \cite{binder81,landau00}, 
and plot the probability distribution of film thicknesses
averaged over lateral areas of increasing size. We consider
$\hh(\rpar)$, which accounts for a lateral size of two unit cells;
$\hh_{1/4}$, which accounts for an average over a quarter of the full 
system, and $\hh$, an average over the full system size. The results
are presented in Fig.~\ref{fig:dist_basal}. 

For the local order parameter $\hh(\rpar)$,
we find rather broad distributions, which span as much as $9$~\AA, i.e.,
$\approx 3$ molecular diameters. 
In the event of a first order phase transition, broad distributions
found in small systems become bimodal, with
two sharp peaks  separated by a gap of increasingly smaller
probability as the system size grows.
Contrary to this scenario, the block analysis
performed over distributions of $\hh_{1/4}$ and $\hh$, 
shows that no signs of bimodality persist in any of these  
distributions, the peaks appear to sharpen significantly, and
the gap between two and three layers is filled with unimodal
distributions. 
However, the distributions corresponding to fully formed
layers are clearly much sharper, while those corresponding
to distances in between remain broader, i.e. exhibit
enhanced fluctuations. 
This scenario resembles that of a continuous phase
transition in a finite system \cite{binder81,landau00}, 
but our block analysis does not 
show evidence of singular behavior. Accordingly, it appears that the
system is traversing the prolongation of a first order
phase transition beyond the critical point, across
a continuous and non-singular transition. This explains
why SFG experiments, which probe a local order parameter,
can detect  distinct spectral changes of the local
environment as temperature is raised.

Indeed, an analysis of average mean square fluctuations and higher moments
(Supplementary Material)
reveals the signature of two rounded layering
transitions on the basal plane, first, from one to two bilayers, 
at about $T$=235~K, then from two to three
bilayers at $T$=267~K. These transitions very much correlate with the inflection
points observed previously in the density profiles. The latter
is consistent with results  by Sanchez et al. for
the TIP4P/Ice model, which  bracketed the bilayer melting 
between 260 and 270~K \cite{sanchez17}. 
   Hints of an inflection in the premelting film thickness
   that could be consistent with the same phenomena are
   also observable in the mW model at about 260~K \cite{pickering18,qiu18}.
For the primary prism plane (Supplementary Material)
the results are similar and reveal the presence of two
rounded transitions at  about 250~K and 267~K, as suggested
from the study of inflection points in the density profile.
The transitions at 235~K and 250~K observed for basal and prism planes
correlate with the onset of surface mobility reported recently for the
same model \cite{kling18}.

Interestingly, the distance between maxima of the global
order parameter distribution increases with film height. 
For the basal plane the three maxima are separated by 2.9 and 3.4~\AA,
respectively. For the prism plane,  the
separation is 2.3, 2.7 and 3.6~\AA, respectively. This is
again consistent with the renormalization scenario, since
we expect the renormalization of  $k_z$ to be more
significant as the distance from the substrate
increases. Thus, it might be possible that true first order transitions 
occur for large $\hh$ between adjacent minima that are
several molecular diameters apart as a result of renormalization.
This view provides an explanation for the experimental
observation of frustrated complete wetting in confocal
microscopy experiments \cite{murata16}.

In summary, we document the existence of rounded layering transitions
with enhanced fluctuations at  235 and 267~K for
the basal interface and at 250 and 267~K for the prism interface. 
We conjecture that the continuation of a line of true first
order layering transitions beyond the layering critical point
could intersect the sublimation line and explain the enhanced 
fluctuations observed here.
This reconciles conflicting interpretations of layering on 
the ice surface \cite{sanchez17,smit17,qiu18}, and opens exciting new avenues 
for experimental verification.

\begin{acknowledgments}
We would like to acknowlege Enrique Lomba for helpful discussions and
Jose Luis F. Abascal for support.
We acknowledge use of the Mare-Nostrum supercomputer and the technical support
provided by Barcelona Supercomputing Center from the Spanish Network of
Supercomputing (RES) under grants QCM-2017-2-0008 and QCM-2017-3-0034. We
also acknowlege funding from Agencia Estatal de Investigaci\'on under 
research grant FIS-89361-C3-2-P.
\end{acknowledgments}


%

\clearpage

\onecolumngrid

\setcounter{page}{1}
\setcounter{table}{0}
\setcounter{figure}{0}
\pagenumbering{arabic}

{\centering
{\large Supporting Information for}

{\Large Rounded layering transitions on the surface of ice 
\\ by \\}
{\large Pablo Llombart$^{\dag,\ddag}$, Eva G. Noya$^\ddag$, David N. Sibley$^\S$, Andrew J. Archer$^\S$ and Luis G. MacDowell$^\dag$}

{\normalsize
$^\dag$Departamento de Qu\'{\i}mica-F\'{\i}sica (Unidad de I+D+i Asociada al CSIC), Facultad de Ciencias Qu\'{\i}micas, Universidad Complutense de Madrid, 28040 Madrid, Spain, \\
$^\ddag$Instituto de Qu\'{\i}mica F\'{\i}sica Rocasolano, CSIC, Calle Serrano 119, 28006 Madrid, Spain and \\
$^\S$Department of Mathematical Sciences, Loughborough University, Loughborough LE11 3TU, United Kingdom.}

}

\vspace*{1cm}

\section*{Results for the Primary Prism facet}

\begin{figure*}[htb!]
   \subfigure[]{\label{perfil_liq1_pI}
   \includegraphics[width=0.3\textwidth,height=0.35\paperwidth,keepaspectratio]{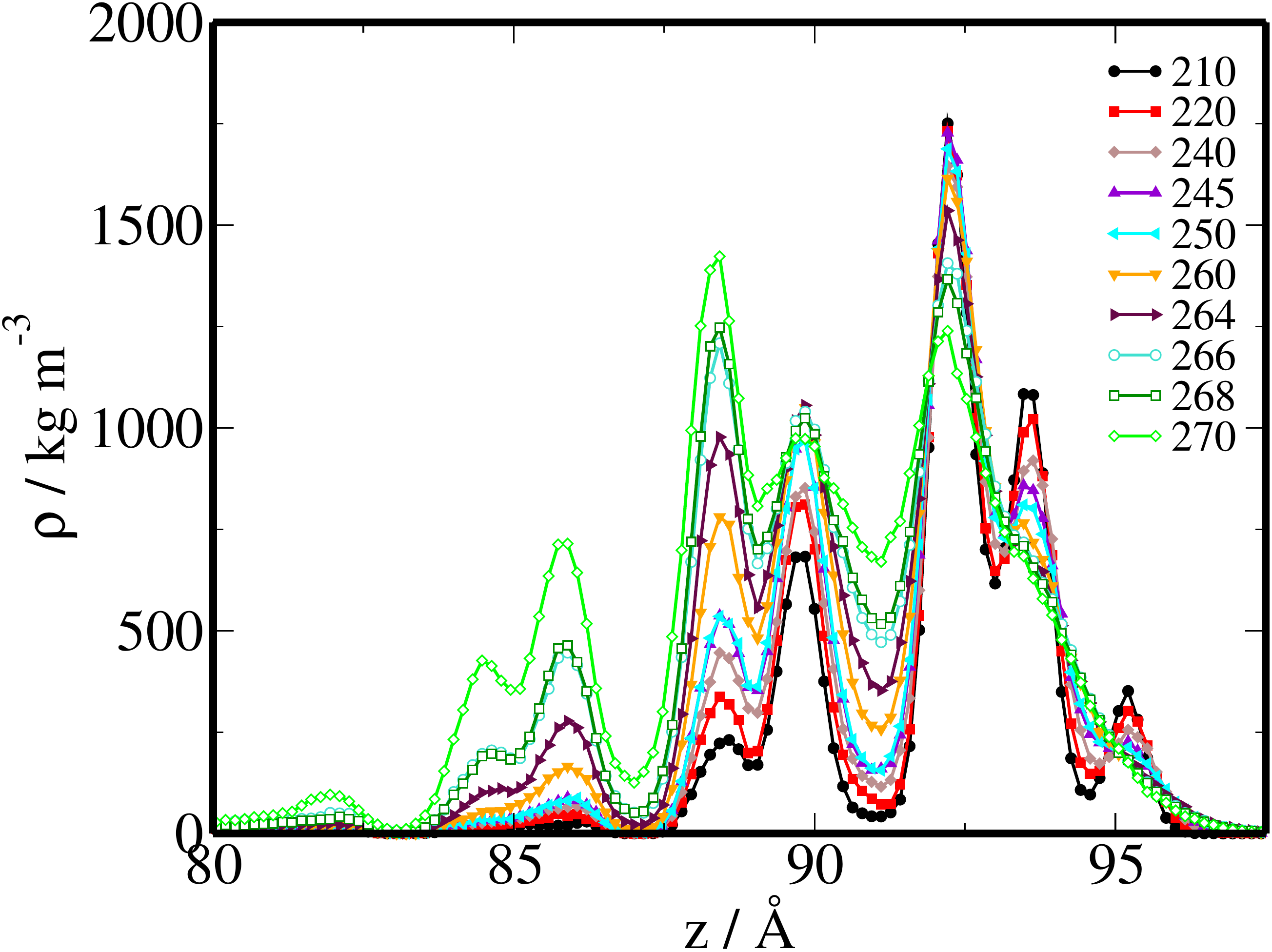}}
   \subfigure[]{\label{perfil_liq2_pI}
   \includegraphics[width=0.3\textwidth,height=0.35\paperwidth,keepaspectratio]{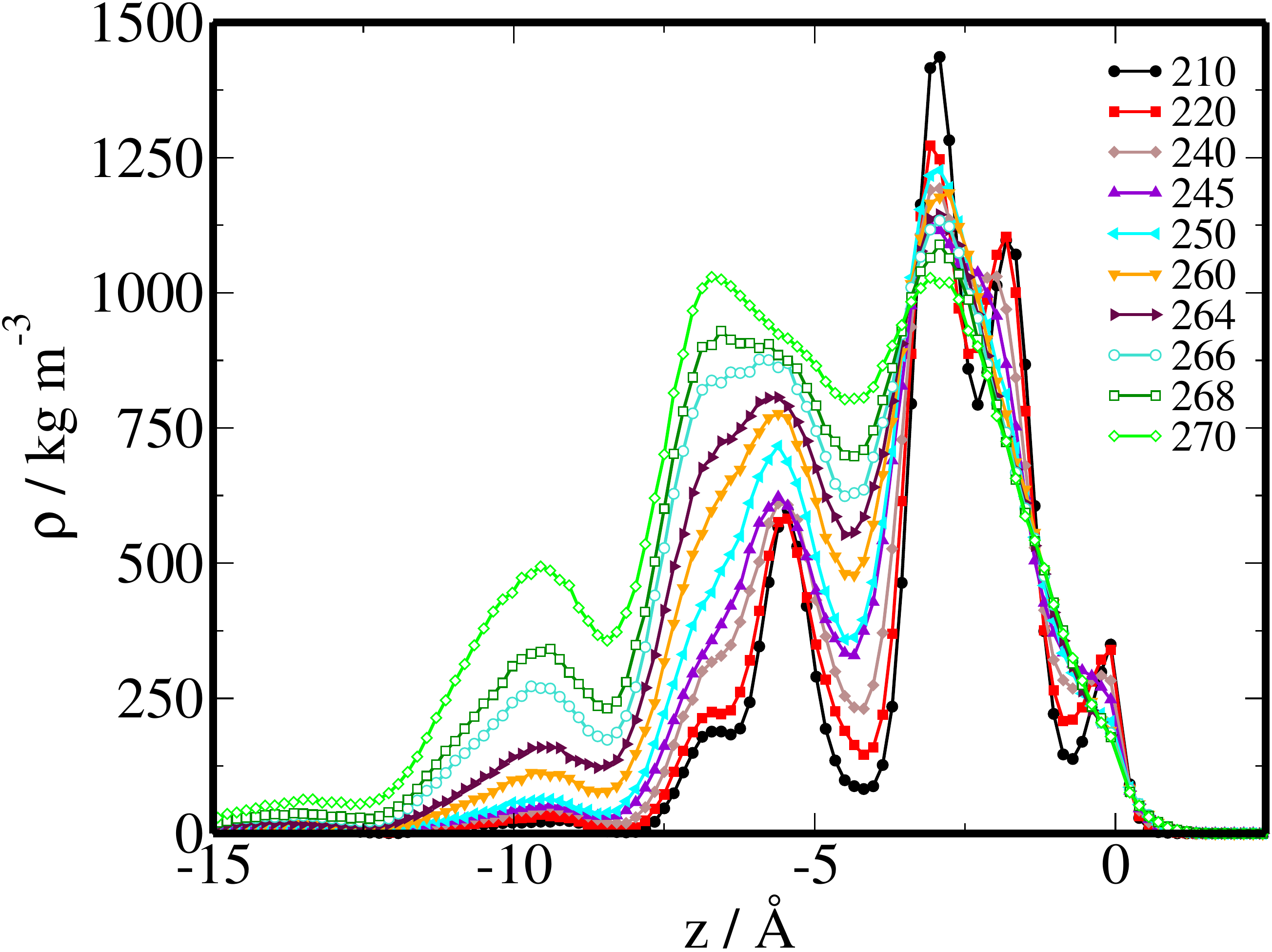}}
   \subfigure[]{\label{perfil_sol2_pI}
	  \includegraphics[width=0.3\textwidth,height=0.35\paperwidth,keepaspectratio]{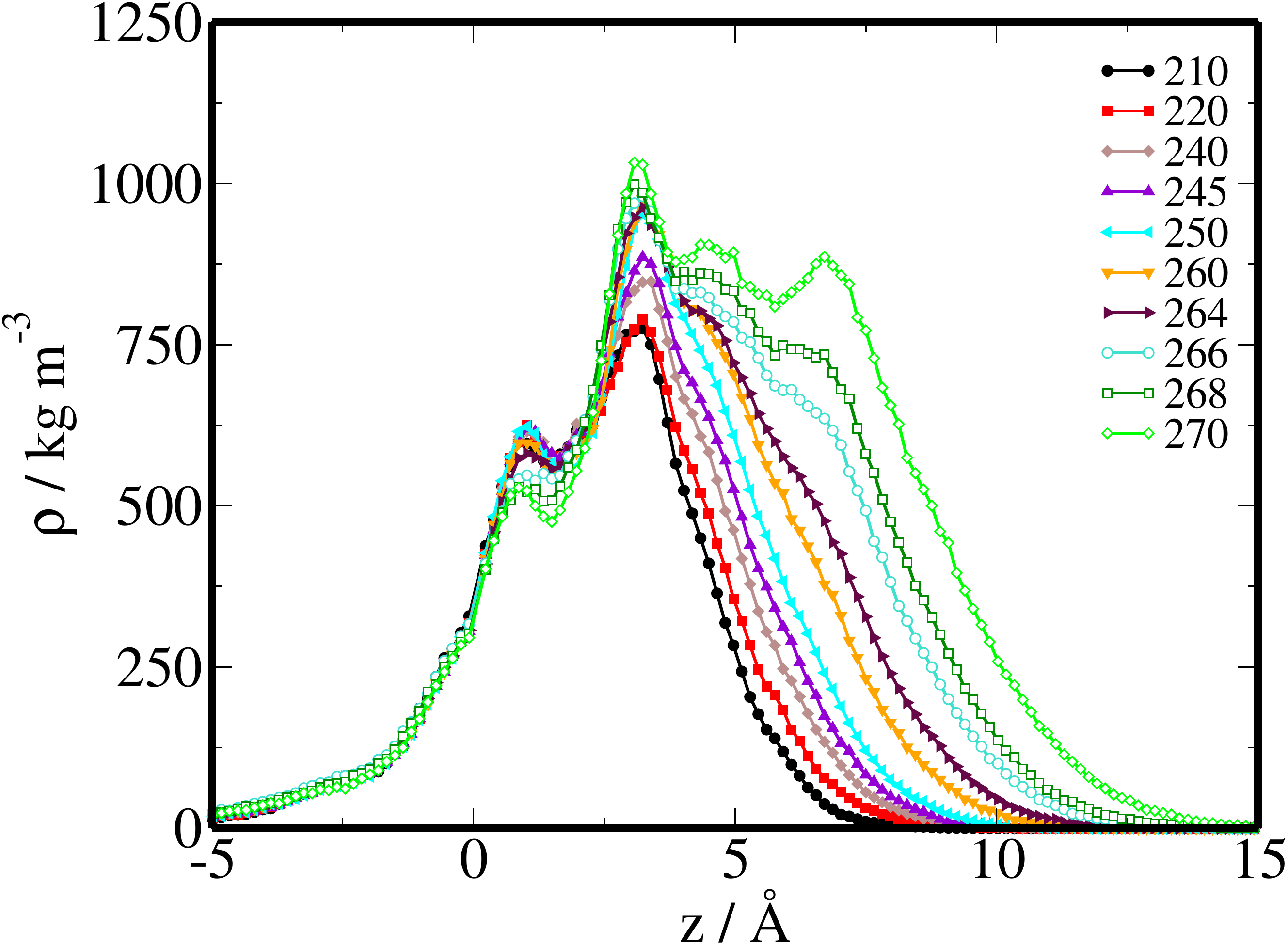}}
   \caption{Density profiles of liquid-like molecules for the
primary prism interface as measured relative to (a) the laboratory reference frame
   (b)   the \qLV~surface and 
 (c)  the \SqL~surface. 
}
	 \label{perfiles_pI}
 \end{figure*}

\begin{figure*}[htb!]
\subfigure[]{\label{fig:espesor_big_pI}
\centering
        \includegraphics[width=0.25\paperwidth,keepaspectratio]{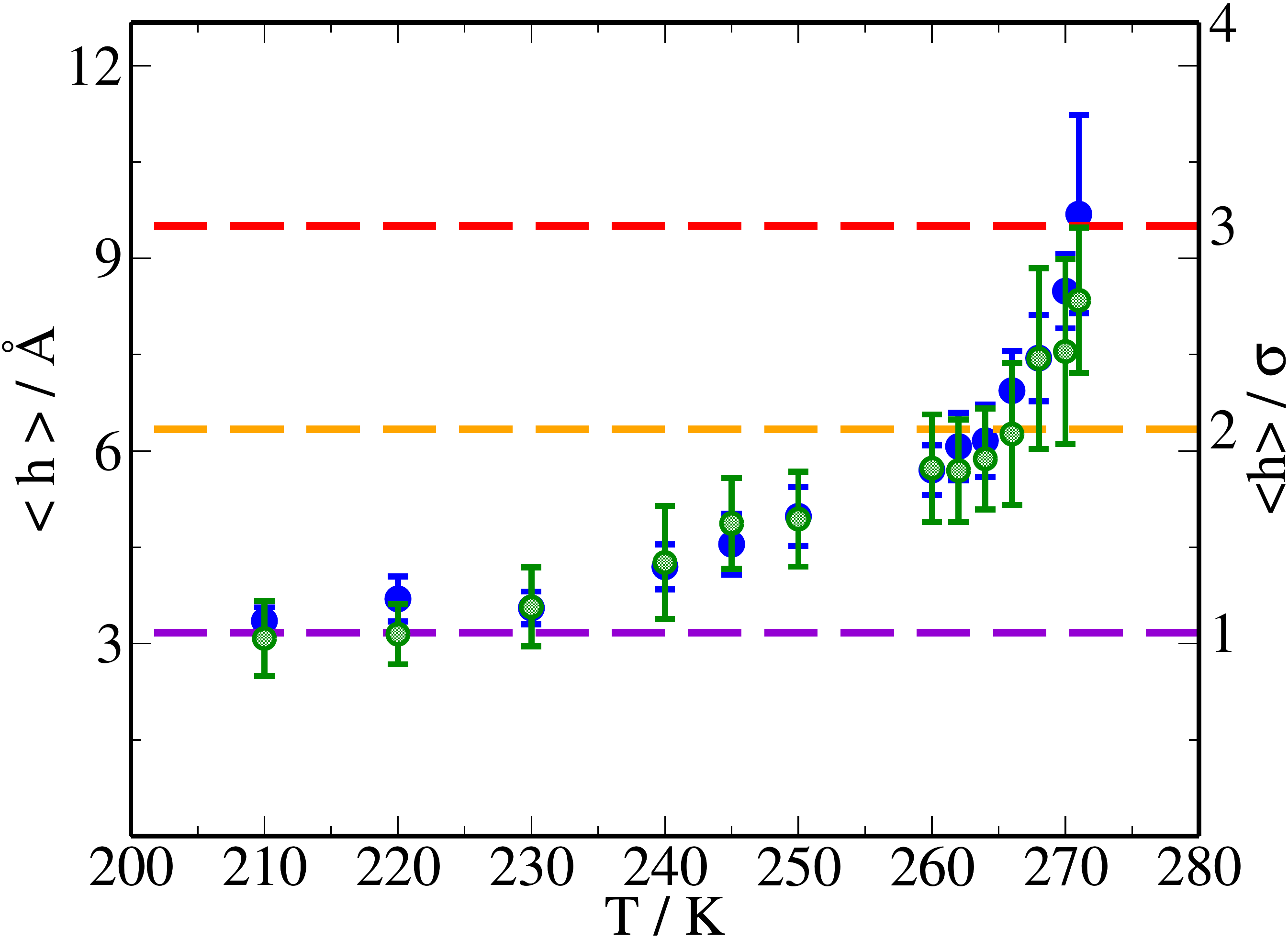}}
 \subfigure[]{\label{fig:disjoining_pI}
\includegraphics[width=0.25\paperwidth,keepaspectratio]{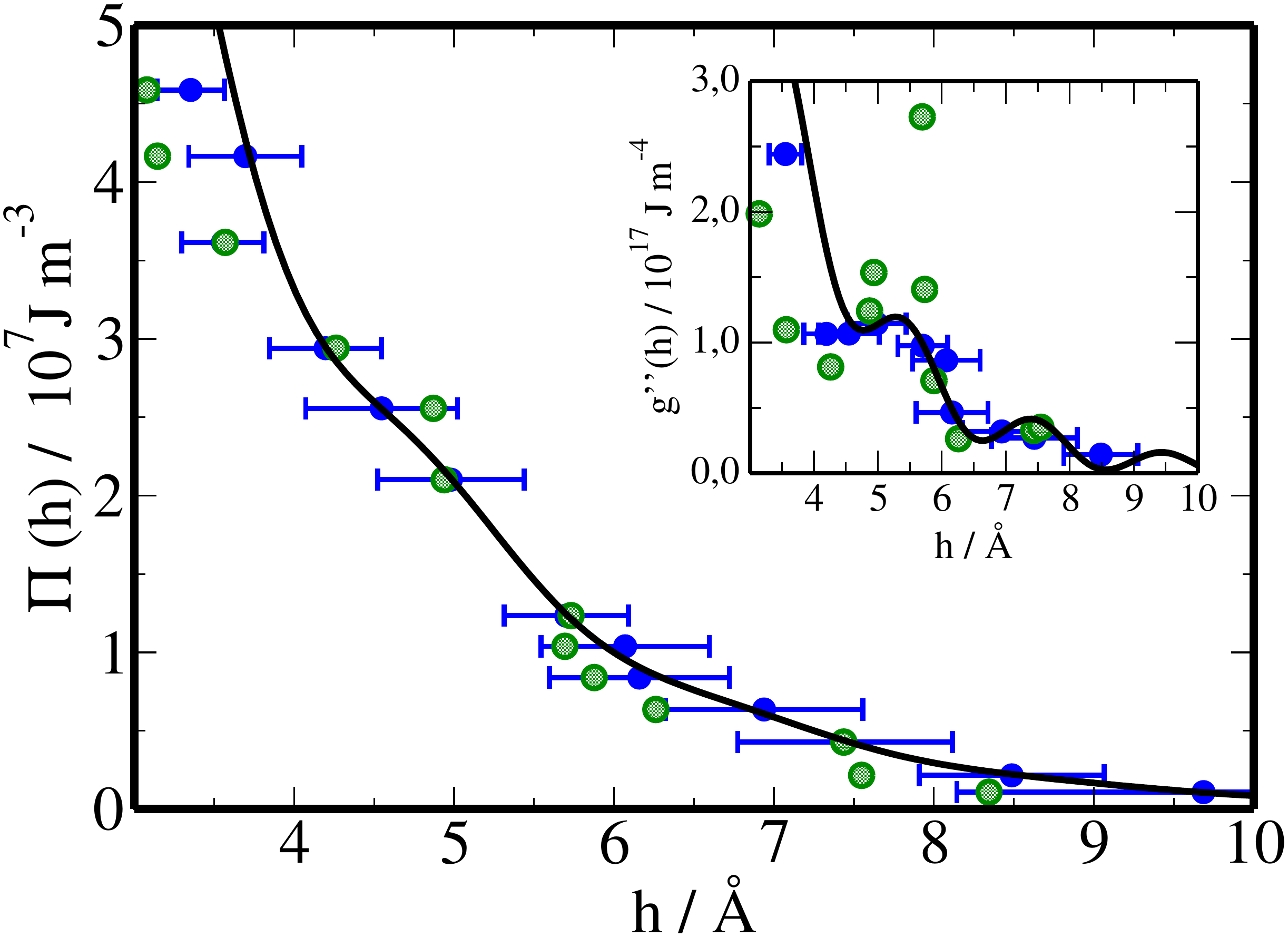}}
\subfigure[]{\label{fig:g_de_h_p}
\centering
      \includegraphics[width=0.23\paperwidth,height=0.18\paperwidth]{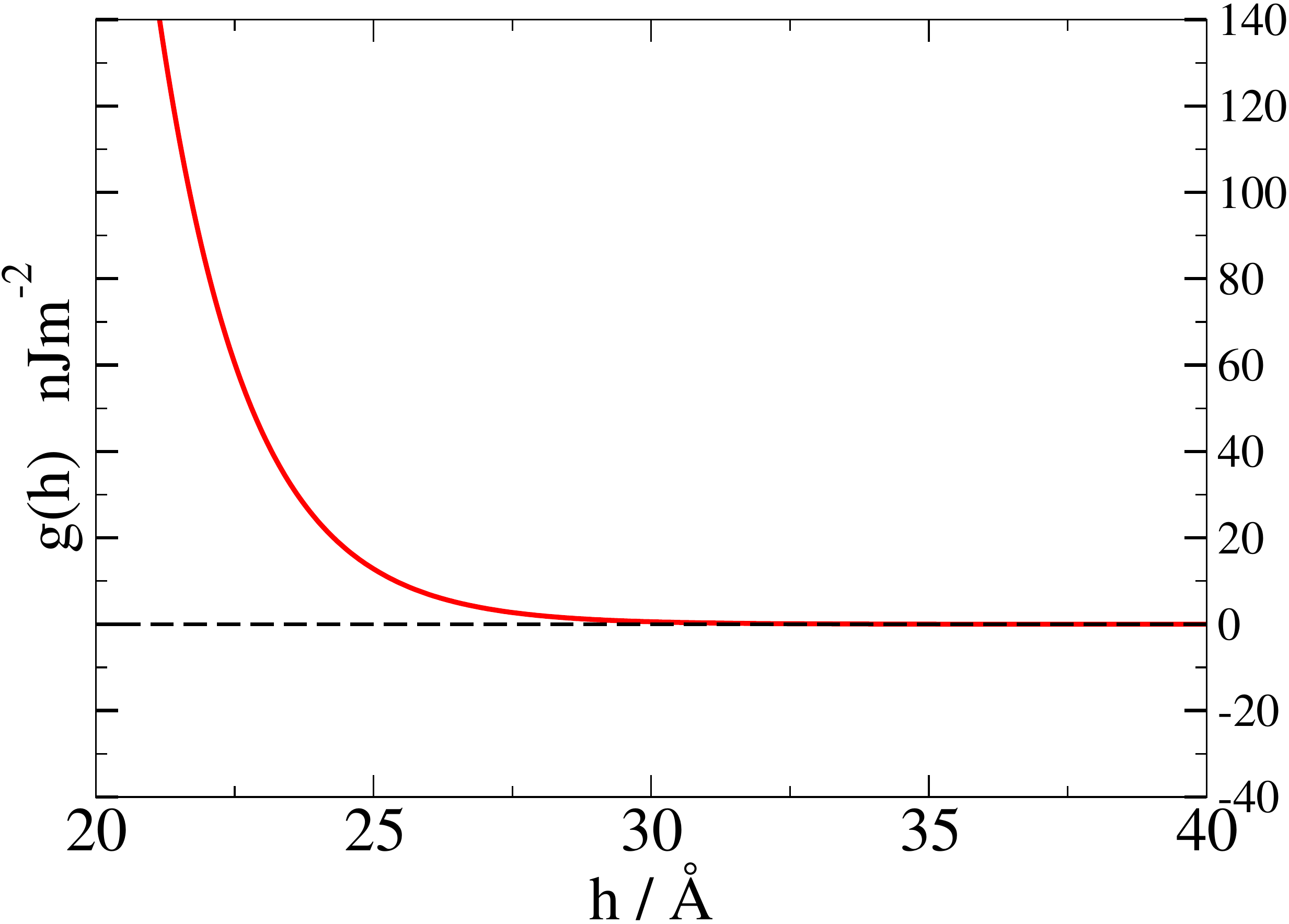}}
	\caption{Film thickness (a), disjoining pressure (b), and interface
potential (c) for the primary prism facet.
      Results are shown for two system sizes with 
      $n_{x}\times n_{y} =5120$ (blue filled circles) and  
      $n_{x}\times n_{y} =1280$ (green hollow circles) molecules.  
	(a) The dashed lines indicate discrete film heights in units 
	of the molecular diameter.  (b) 
        The lines are a fit to \Eq{grn}. The insets show inverse surface
	  susceptibilities as obtained from numerical derivatives (symbols) and
	  from analytical derivatives of the fits (lines). Error bars here are
      shown only for the data fitted to \Eq{grn}.
       Panel (c) shows the decay of the interface potential at 
intermediate distances as extrapolated from the fit to \Eq{grn}. 
The decay of the interface potential here is monotonic.
   }
	\label{espesores_pI}
 \end{figure*}

\begin{figure*}[htb!]

  \subfigure[]{\label{p_global_p}
     \includegraphics[width=0.3\textwidth,height=0.35\textwidth,keepaspectratio]{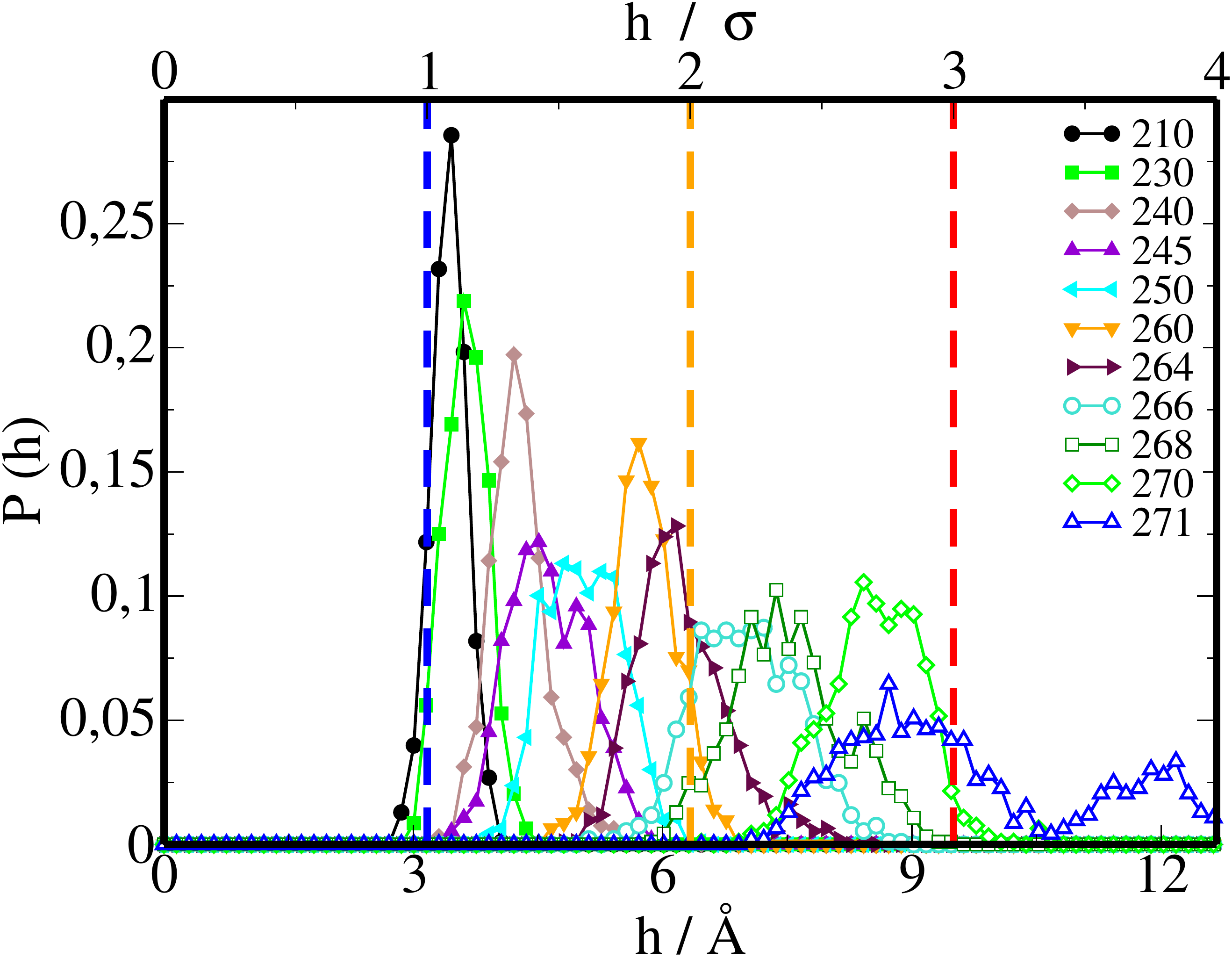}}
  \subfigure[]{\label{p_semilocal_p}
     \includegraphics[width=0.3\textwidth,height=0.35\textwidth,keepaspectratio]{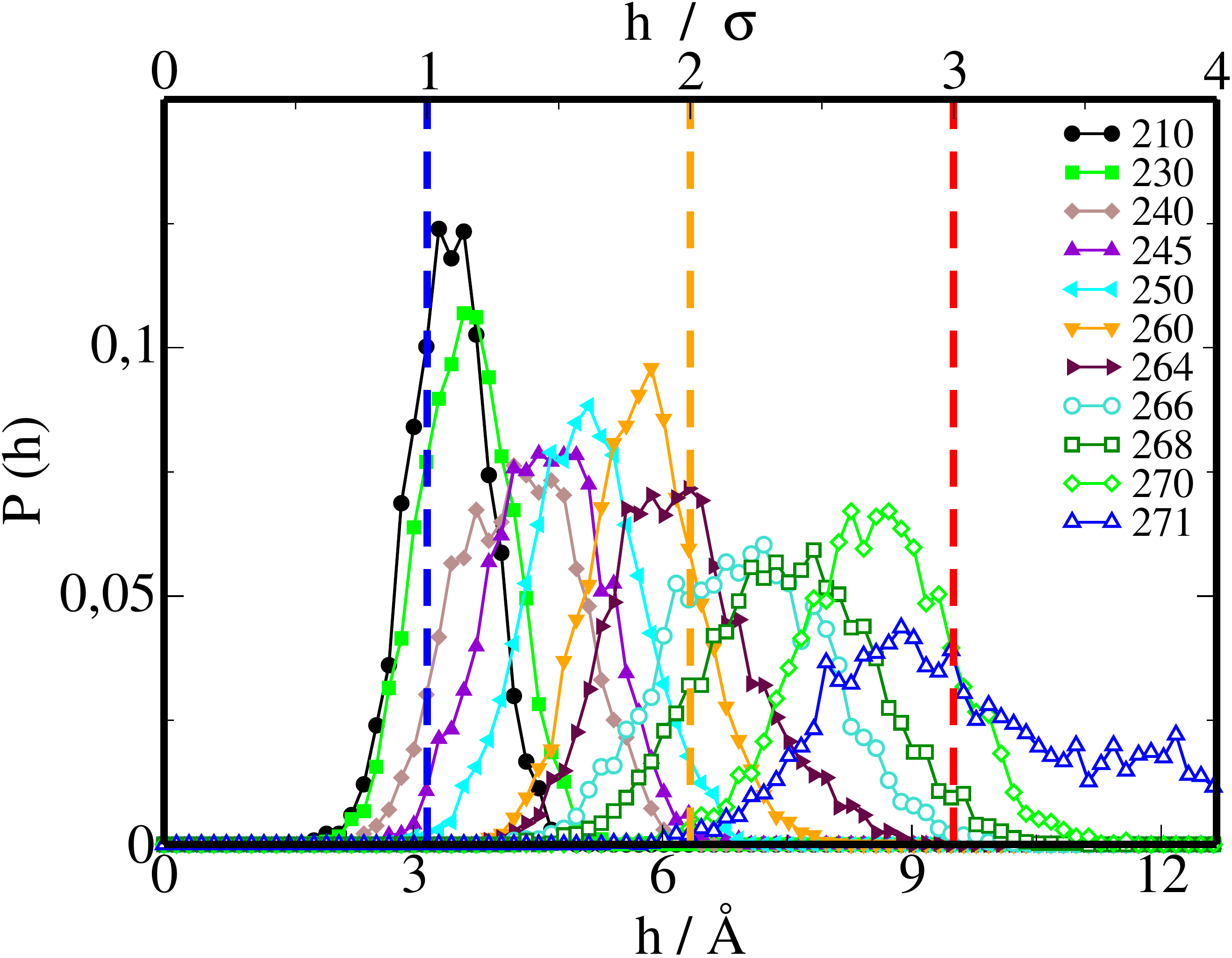}}
  \subfigure[]{\label{p_local_p}
     \includegraphics[width=0.3\textwidth,height=0.35\textwidth,keepaspectratio]{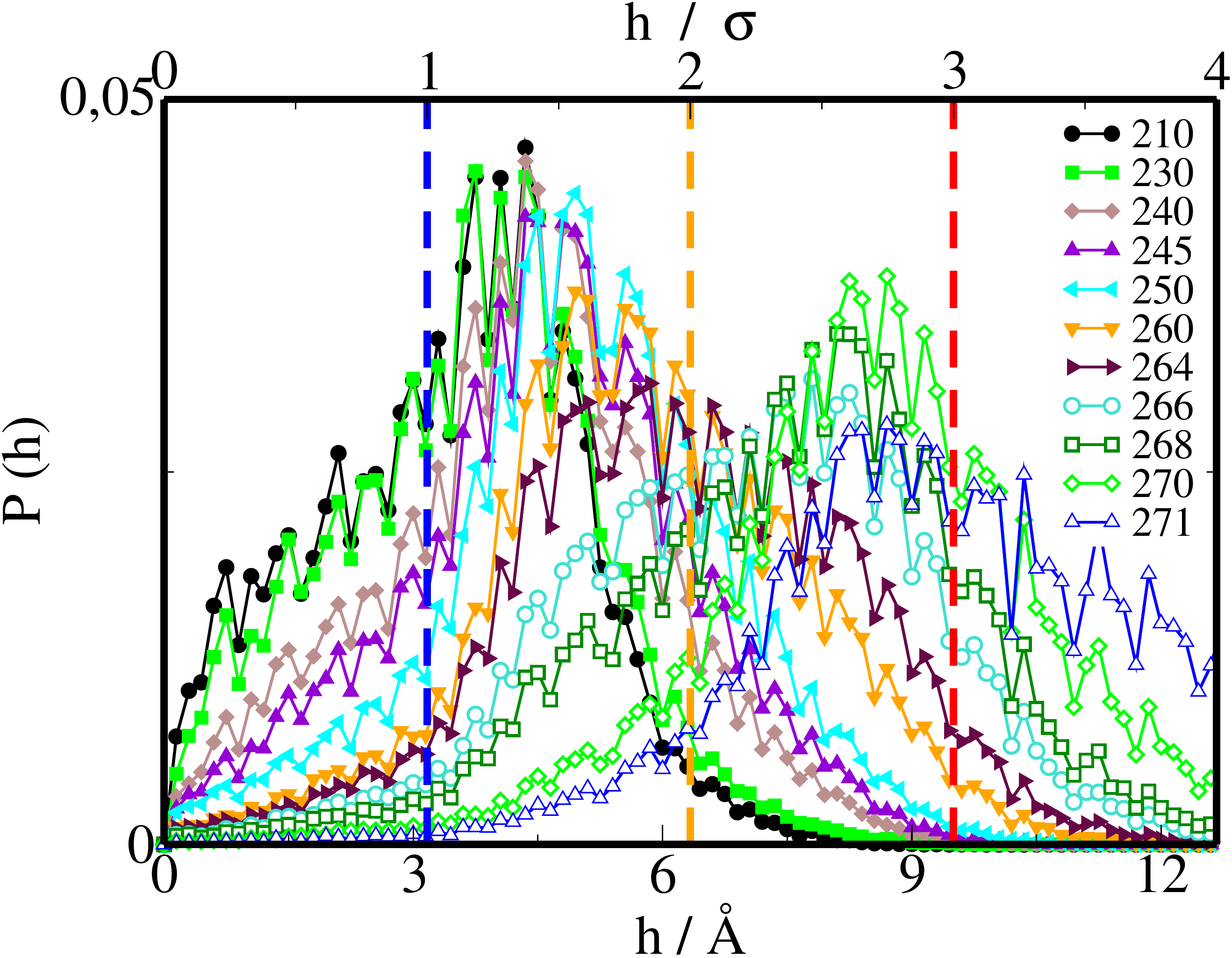}}
	\caption{System size analysis of thickness distributions at the
	primary prism plane.  Panels (a) probability distribution of the
global film thickness, $h$. (b) probability distribution of the partial
film thickness $h_{1/4}$. (c)  probability distribution of the local order
parameter $h(\rpar)$. Dashed vertical lines show the film thickness in units of
the molecular diameter. Results are shown for temperatures in the range from
$T$=210 K to $T$=271 K, with color code as indicated on the figure. 
 }
 \label{fig:dist_pI}
\end{figure*}

\section*{Disjoining pressure}

The three phases involved are
the ice substrate ({\it i}), the water film ({\it f}) and the vapor ({\it v}).
The discussion below is rather general, so here we refer
to the three phases as solid ({\it s}), liquid ({\it l}) and vapor ({\it v}).
Consider an inert substrate in contact with a stable vapor phase at fixed 
temperature, $T$.  It is convenient to study the system in the
grand canonical ensemble, where the chemical potential $\mu$ is fixed
at the outset, and one assumes the adsorbed layer can grow to its equilibrium 
value while the vapor pressure remains constant. By this device,
one fixes the vapor pressure of the system to the corresponding bulk
vapor pressure $p_v(T,\mu)$ at the selected chemical potential.
The corresponding bulk liquid phase of equal temperature and chemical
potential has a bulk pressure, $p_l(T,\mu)$.

Consider that $(T,\mu)$ are  such that the bulk vapor phase
is more stable than the bulk liquid phase.  For a  column
of height $\hh$ filled by bulk vapor, the free energy per unit surface
is given by $\omega_v(T,\mu)=-p_v(T,\mu)\hh$.
Similarly, for an equivalent column filled with bulk liquid,
the corresponding free energy 
is given as  $\omega_v(T,\mu)=-p_l(T,\mu)\hh$. Below
the bulk liquid-vapor coexistence, it always holds $p_v(T,\mu) > p_l(T,\mu)$,
such that the vapor phase has the lowest  free energy
and is the preferred phase. The free energy difference per
unit surface area between
a liquid filled and vapor filled column is  given
as $\Delta \omega(T,\mu) = \Delta p\lv(T,\mu) \hh$, where
$\Delta p\lv(\mu,T) =- [p_l(T,\mu) - p_v(T,\mu) ]$ is
the familiar Laplace pressure difference used in classical
nucleation theory.

If the column of bulk phase is moved next to the substrate,
however, a finite amount
of liquid phase is usually stabilized, even if the thermodynamic 
fields have been selected such that the bulk vapor phase 
is most stable. The interface potential, $g(\hh)$ accounts
for the free energy gained by the liquid phase in contact
with the substrate.

The excess free energy of the liquid film of thickness $\hh$ 
adsorbed on a substrate is given as:
\begin{equation}\label{eq:omega}
   \omega(\hh;T,\mu) = \gamma_{sl}+\gamma_{lv} + g(\hh;T) - \Delta p\lv(T,\mu) \hh,
\end{equation} 
Here, $- \Delta p(T,\mu) \hh$,  is the bulk free energy cost to
replace the bulk vapor by a liquid film, $g(\hh)$ accounts
for the free energy gained by the bulk liquid phase due to
the influence of the substrate at liquid-vapor coexistence,
while $\gamma_{sl}$ and $\gamma_{lv}$ are the solid-liquid and
liquid-vapor interfacial tensions, respectively.

The condition of equilibrium for the adsorbed film at constant
$T,\mu$ is:
\begin{equation}
   \frac{d \omega(\hh;T,\mu)}{d \hh} = 0.
\end{equation} 

This provides the celebrated
Derjaguin equation \cite{derjaguin87,henderson05,benet14b}:
\begin{equation}\label{eq:derjaguin2}
   \Pi(\hh;T) = - \Delta p\lv(T,\mu), 
\end{equation}
where $\Pi(\hh;T)$ is the disjoining pressure, defined as
$\Pi(\hh;T) = - d g(\hh;T) / d\hh$.


In standard applications for adsorbed liquids on an inert substrate, one
usually measures an adsorption isotherm, i.e., $\hh=\hh(T,\mu)$,
at fixed temperature, while changing the chemical potential (or vapor pressure).
The disjoining pressure is calculated by matching the measured equilibrium 
film thickness at increasing vapor pressure with the corresponding 
pressure difference, $\Delta p\lv(T,\mu)$ \cite{benet14b}.

For a premelting film, the situation is far more complicated, because
adsorption from the vapor phase takes place on a solid substrate
that is found at phase coexistence with the vapor.  i.e.,
the vapor-substrate equilibrium states only exist along the
sublimation or solid-vapor coexistence line, corresponding
to a fixed chemical potential $\mu\sv=\mu\sv(T)$.
Therefore, at fixed temperature one cannot
meaningfully collect equilibrium film thicknesses $\hh$ as a function
of vapor pressure, since equilibrium is achieved only at exactly
the solid-vapor coexistence pressure, $p\sv(T,\mu\sv(T))$.

The way out is to relax the constraint of fixed temperature.
Instead, we collect the equilibrium film thicknesses
$\hh(T)|\sv=\hh(T,\mu\sv)$ along the sublimation line,
where the notation $X(T)|\sv$  denotes a property $X(T)$ measured at the
chemical potential $\mu\sv=\mu\sv(T)$ of 
solid/vapor coexistence. These film thicknesses are 
then mapped onto $\Delta p\lv(T)|\sv$, i.e., the pressure difference
between bulk liquid and vapor phases  measured
at the solid-vapor coexistence chemical potential. 

   Taking into account that the temperature dependence of
   $g(\hh;T)$ and $\Pi(\hh;T)$ is small, one can get
   a single disjoining pressure master curve $\Pi(\hh)$ from
   these data. Accordingly, the function $\Pi(\hh)$ is obtained in three steps:
\begin{itemize}
\item Measure the equilibrium film thickness $\hh(T)|\sv$ along the
	sublimation line. In practice, this is readily done by direct coexistence
	$NVT$ simulations. Placing a slab of ice in contact with vacuum
	at a temperature $T$, the system gradually equilibrates and attains
	solid-vapor coexistence. Accordingly, once the system is well
	equilibrated, the vapor phase is automatically
	selected to exhibit the sublimation vapor pressure without
	a priori knowledge of the corresponding chemical potential. 
	 In between the almost empty vapor phase
	and the bulk ice phase, a premelting liquid film is formed in
	a few nanoseconds. 
	The  equilibrium film thickness that is obtained corresponds
	to the desired curve, $\hh(T)|\sv$.
\item Calculate the Laplace pressure difference between bulk liquid and vapor
   phases at the solid-vapor coexistence chemical potential,
   $\Delta p\lv(T)|\sv$. Below we show that this pressure difference
   can be related to the liquid-vapor and solid-vapor coexistence pressures.
   Using the Clapeyron equation the coexistence pressures
   may be obtained from triple point data only. In practice,
   we assume an ideal gas for the vapor phase, and obtain accurate
   approximations for the solid and liquid properties at the triple point
   from $NpT$ simulations at zero pressure (see below).
\item Map the results of $\hh(T)|\sv$ onto $\Delta p\lv(T)|\sv$ to
   obtain the disjoining pressure curve  from \Eq{derjaguin2} as:
   \begin{equation}
   \Pi(\hh(T)|\sv)  = - \Delta p\lv(T)|\sv
   \end{equation} 
\end{itemize} 


\section*{Laplace pressure difference along the sublimation line}

In order to evaluate the disjoining pressure, we need to know
the Laplace pressure difference $\Delta p\lv(T)|\sv$ between
the bulk liquid and vapor phases at equal temperature and chemical potential,
for selected chemical potentials along the sublimation line.

It turns out that this calculation can be performed 
following thermodynamic calculations that are familiar in
classical nucleation theory.

Here, we are interested in calculating the pressure difference
$\Delta p\lv(T) = p_l(T,\mu) - p_v(T,\mu)$ between an adsorbed
liquid phase and a reservoir vapor phase with chemical potential $\mu$.
Using the Gibbs-Duhem equation, $dp = \rho\, d\mu$, with $\rho$, the
bulk density, we write:
\begin{equation}
  d (p_l - p_v) = (\rho_l - \rho_v) d\mu.
\end{equation} 

Since the vapor density is much smaller than the liquid density,
we assume  $\rho_l - \rho_v \approx \rho_l$. Furthermore, since
the liquid is effectively incompressible for the small
chemical potential changes that concern us here, we assume
constant liquid density and
integrate the above equation from the liquid-vapor coexistence
chemical potential to arbitrary chemical potential at
fixed temperature, yielding:
\begin{equation}
   p_l(T,\mu) - p_v(T,\mu) = \rho_l ( \mu - \mu\lv),
\end{equation} 
where $\mu\lv$ is the chemical potential at liquid-vapor coexistence
and we have taken into account that  $p_l(T,\mu\lv)- p_v(T,\mu\lv)= 0$.

Now, assuming the vapor phase is an ideal gas, we obtain:
\begin{equation}
   p_l(T,\mu) - p_v(T,\mu) = \rho_l  RT \ln p/p\lv(T),
\end{equation} 
where $p(T)\lv$ is the liquid-vapor coexistence pressure
and $p$ is the pressure of the vapor phase with arbitrary
chemical potential $\mu$.
This is the result used in classical nucleation theory
for the pressure difference of a liquid nucleus in equilibrium
with the vapor reservoir phase. Here we use the same equation
for the adsorbed liquid layer in equilibrium with a vapor
reservoir phase.

It follows immediately that the sought Laplace pressure
difference along the sublimation line, where $p=p\sv$ by
definition, is given as:
\begin{equation}
   \Delta p\lv|\sv = \rho_l  RT \ln\frac{p\sv(T)}{p\lv(T)}.
\end{equation} 


Therefore, $\Delta p\lv(T)|\sv$ may be calculated from knowledge of
the vapor pressures along the condensation and sublimation lines.

In order to obtain $p\sv(T)$ and $p\lv(T)$, we invoke
the Clausius equation for the coexistence pressure, $p_{\alpha v}$ 
of an arbitrary phase, $\alpha$, with the bulk vapor:
\begin{equation}
   \frac{d p_{\alpha v}}{d T} = \frac{\Delta H_{\alpha v}}{T\Delta V_{\alpha v}},
\end{equation} 
where $\Delta H_{\alpha v}$ and $\Delta V_{\alpha v}$ are the enthalpy
and volume change of transition. For water below the triple point,
the volume of the vapor phase is more than 1000 times larger than that
of either liquid or solid phases, so we can approximate
$\Delta V_{\alpha v}$ to the volume of the vapor phase $V_v$, and further
replace this with the ideal gas equation, leading to the
Clausius-Clapeyron result:
\begin{equation}
   \frac{d p_{\alpha v}}{p} = \frac{\Delta H_{\alpha v}}{R T^2} dT.
\end{equation} 

In many applications, this equation is integrated under the assumption
of constant $\Delta H_{\alpha v}$. We improve this approximation
by expanding the enthalpy of phase change to first order about
the triple point:
\begin{equation}
   \Delta H = \Delta H_t + \Delta C_{p,t} (T - T_t),
\end{equation}
where $\Delta C_{p,t}$ is the difference of heat capacities between
the corresponding condensed phase and the coexisting vapor phase
at the triple point.
This then finally yields:
\begin{equation}\label{eq:pvapgen}
   \ln (p_{\alpha v}(T)/p_t )=
   \frac{ \Delta H_t - \Delta C_{p,t} T_t}{R}
       \left [ 1/T_t - 1/T \right ] + \frac{\Delta C_{p,t}}{R}\ln (T/T_t),
\end{equation}
where here $\Delta H_t$ is the enthalpy of phase change, and
the subindex $\alpha$ corresponds to either liquid or solid phases.

In order to evaluate this equation, we calculated enthalpies
and heat capacities for the TIP4P/Ice model at the triple point
temperature $T_t=272$~K and zero pressure. Results for the required
properties may be found in Table \ref{datos_entalpias}.

\begin{table}[htb!]
 \footnotesize
  \begin{tabular}{c c c c}
  \hline
  \hline
  $H^{iv}_t$ / KJ mol$^{-1}$ & $H^{wv}_t$ / KJ mol$^{-1}$ & $C_{p,t}^{i}$/ J
  mol$^{-1}$K$^{-1}$ & $C_{p,t}^{w}$ / J mol$^{-1}$K$^{-1}$ \\ 
  \hline
     -61.03                      &      -55.64                 &      65.60
&                  107.26      \\
  \hline
  \hline
 \end{tabular}
 \caption{Thermophysical properties of the TIP4P$/$Ice model at the triple
 point $T_t=272$~K, required for the determination of the sublimation and
 condensation pressures.}
 \label{datos_entalpias}
\end{table}


\section{Moments of the distributions and rounded transitions}

\begin{figure*}[htb!]
  \subfigure[]{\label{pichi1}
     \includegraphics[width=0.45\textwidth,height=0.45\textwidth,keepaspectratio]{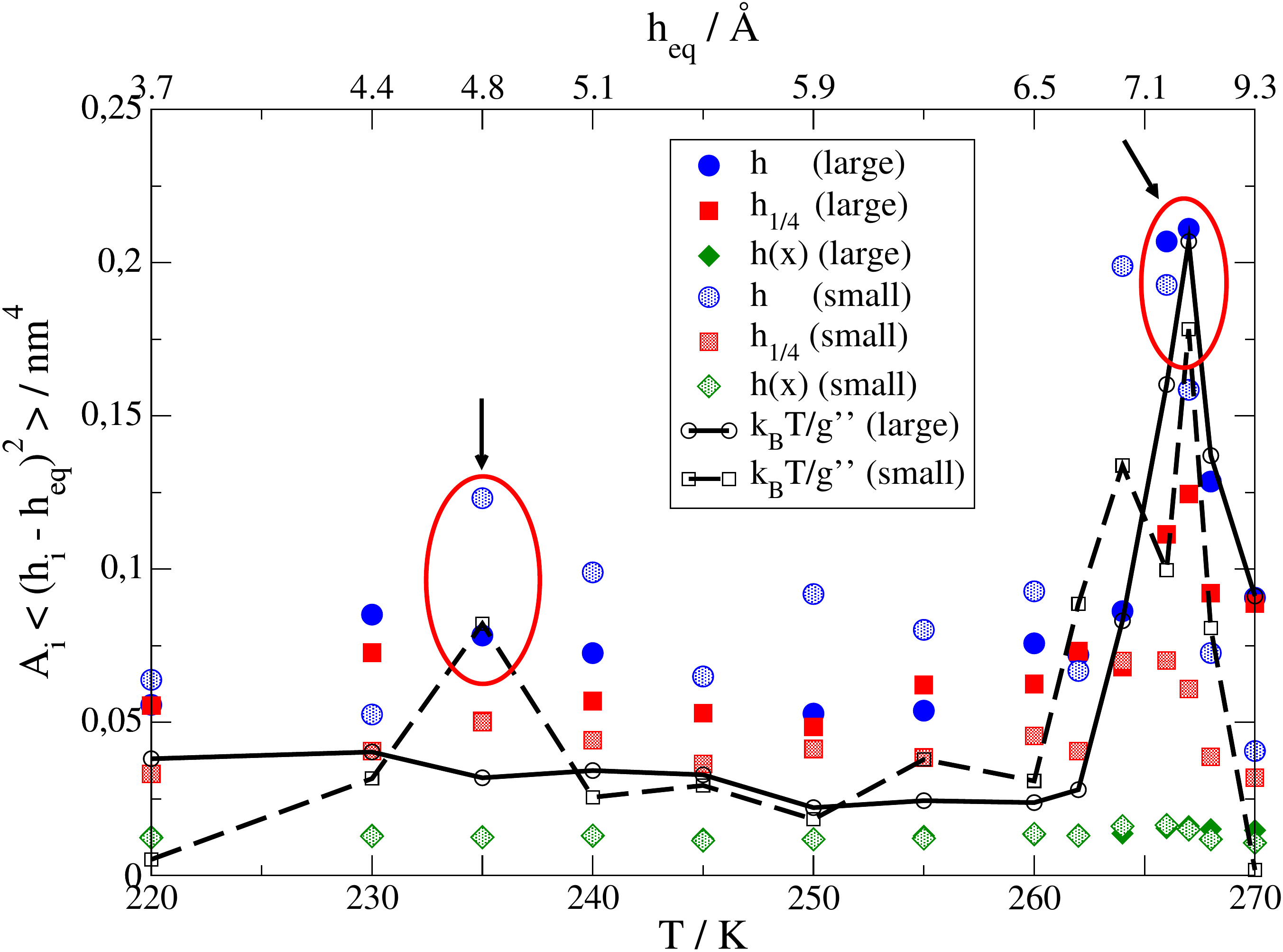}}
  \subfigure[]{\label{pichi2}
     \includegraphics[width=0.45\textwidth,height=0.45\textwidth,keepaspectratio]{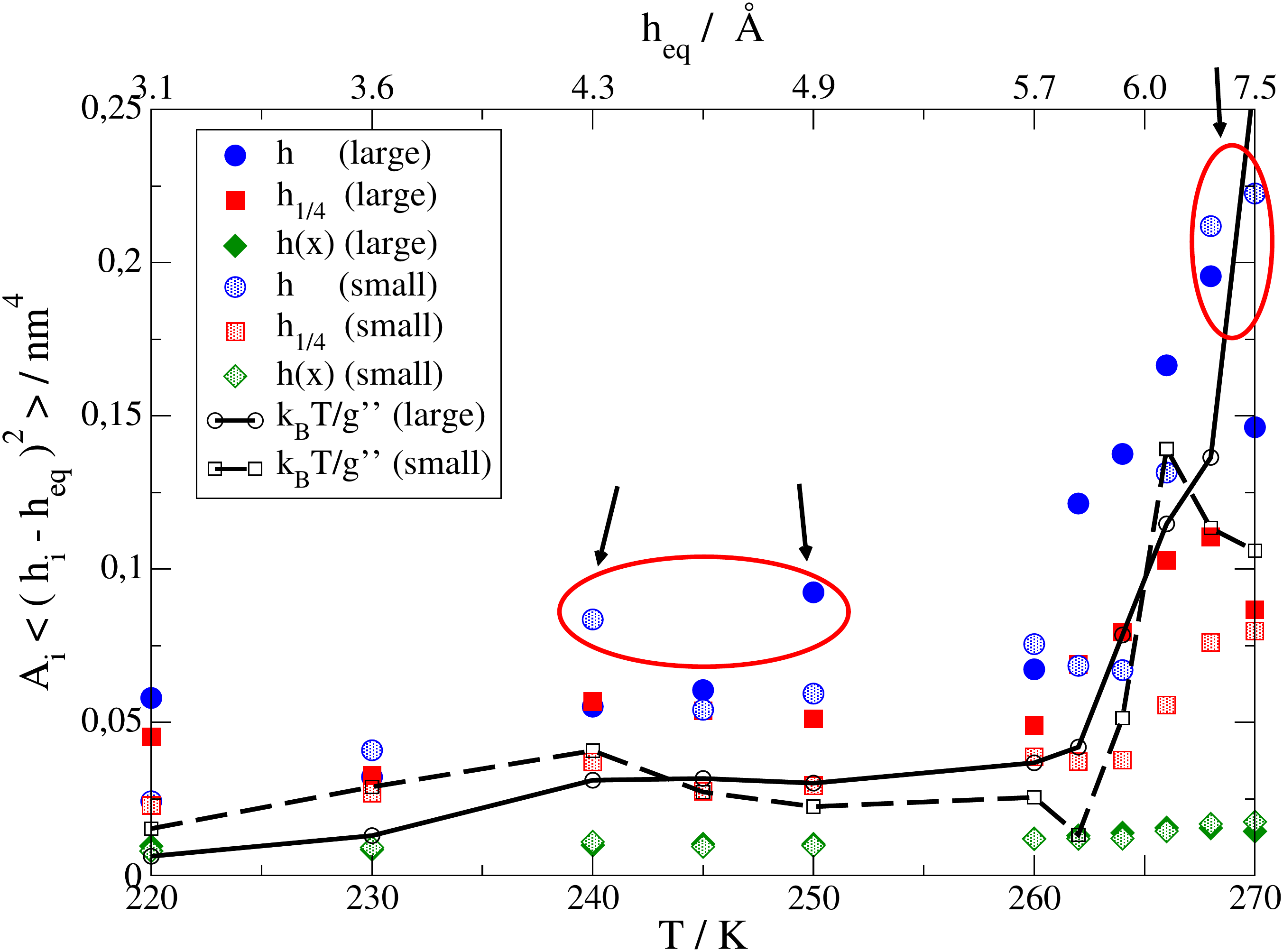}}
	\caption{
Scaled mean squared fluctuations of the film thickness. Results
are shown for the large (filled symbols) and small (hollow symbols) systems. The
latter is  $2\times 2$  smaller than the former. The figure shows moments for
fluctuations of the global film thickness, $\hh$ (circles),  for subsystems with
size 1/4, $h_{1/4}$ (squares) and for the local thickness $\hh(x)$ (diamonds).
Black lines show the scaled mean squared fluctuations as given by the
right hand side of \Eq{msf}, with  $\chi_{\parallel}^{-1}$ estimated from
the second derivatives of the interface potential,  $g''(\hh)$. Results are
shown  for large (full lines) and small (dashed lines) systems. Arrows point roughly to the
location of the maxima of mean squared fluctuations.
 \label{fig:dqm2}
 }
\end{figure*}

\begin{figure*}[htb!]
     \includegraphics[width=0.5\textwidth,height=0.5\textwidth,keepaspectratio]{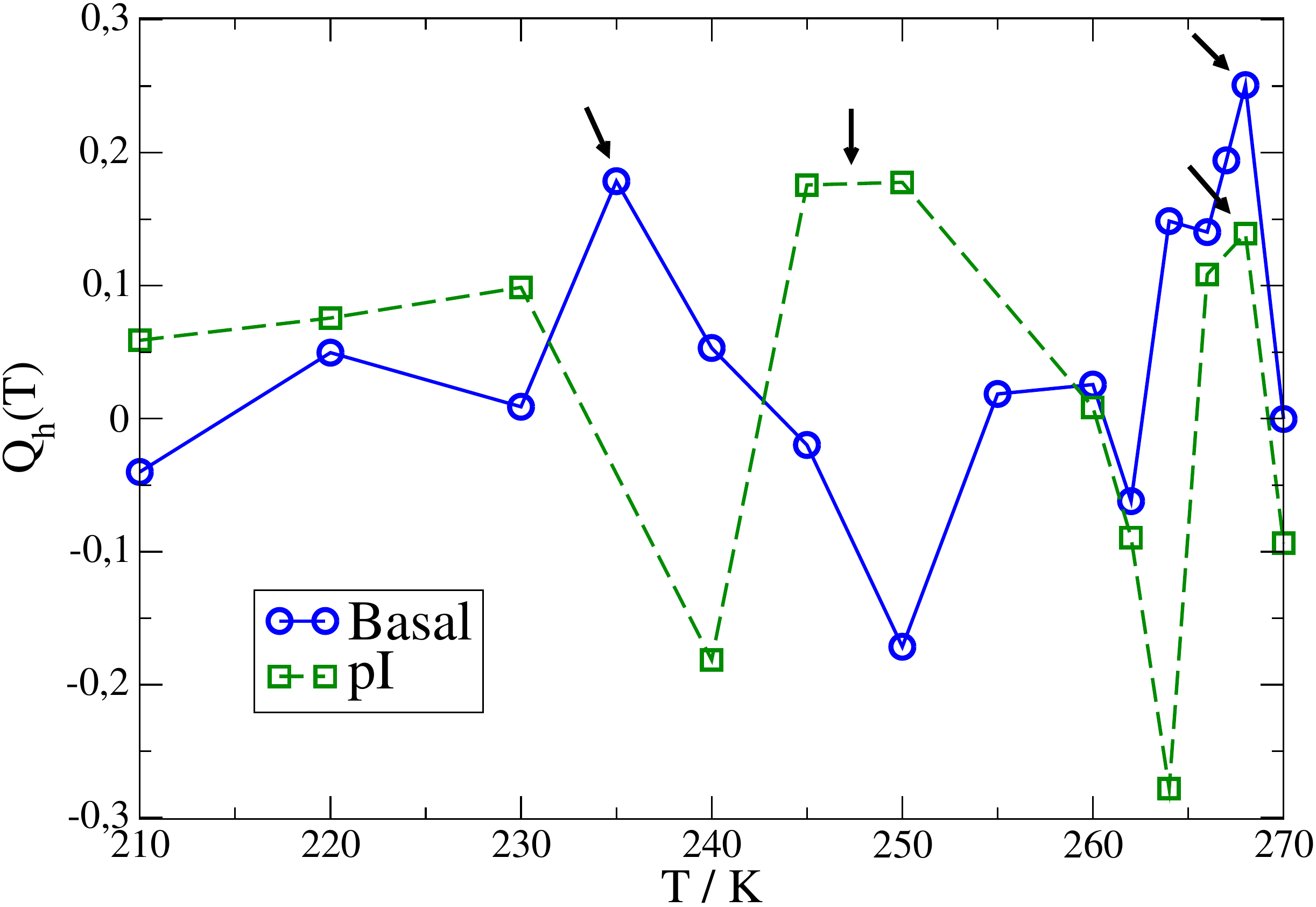}
	\caption{Binder cumulants of the global order parameter for the large
(blue circles) and small (green squares) systems. The arrows point to maxima of
the cumulants, which serves to roughly locate the transitions. 
 \label{fig:cumulants}
 }
\end{figure*}
Since the probability distributions of the order parameter in Fig.~3
of the main manuscript and Fig.~3 of the supplementary material are normalized
to unity, a rough estimate of the mean squared amplitude of the
fluctuations can
be made from the height of the maxima of the distributions. From these plots, one finds that
the maxima corresponding to partially filled layers are less pronounced than
those corresponding to filled  layers. Accordingly, the fluctuations are larger for the former
and we can estimate the loci of the rounded transitions from the distribution
with largest fluctuations.  By visual inspection, this suggests rounded
transitions at about 235~K and 266~K for the basal face,
and at about 250~K and 266~K for the prism face. Of course, because of the
limited statistics, the finite size transitions are given with an error of
$\pm 2$~K at least.

A quantitative analysis requires consideration of the probability distribution
in more detail. In our simulations, the system is found at solid vapor
coexistence by construction. The chemical potential is therefore fixed
exactly at coexistence. Furthermore, the system features a large bulk
phase, which roughly serves as a particle and heat reservoir for the premelting
film. Therefore, we may consider the premelting film as effective being treated
grand canonically.
For a system away from the line of layering transitions,
the surface free energy in \Eq{omega} at the imposed constant temperature and chemical potential
can be expanded to quadratic order as:
\begin{equation}
  \omega(\hh;T,\mu) = \omega(\hh_{eq},T,\mu) + \frac{1}{2} \chi_{\parallel}^{-1} (\hh - \hh_{eq})^2,
\end{equation} 
where $\hh_{eq}$ is the equilibrium film thickness, and the susceptibility
has been defined as $ \chi_{\parallel}^{-1}=g''(\hh_{eq})$. The surface free
energy is related to the probability distribution as $A\omega(\hh;T,\mu)=-k_B T
\ln P(\hh;T,\mu)$, where $A$ is the lateral area. Accordingly, the probability
distribution of the order parameter is (c.f.  Ref.\cite{binder81,landau00}):
\begin{equation}
  P(\hh) = C e^{-\frac{1}{2} \beta A \chi_{\parallel}^{-1} (\hh - \hh_{eq})^2 },
\end{equation} 
with $C$ a normalization constant. This has a Gaussian form, so 
the mean squared fluctuations of $\hh$ in a system with lateral size $A$ will
obey:
\begin{equation}\label{eq:msf}
   A \langle ( \hh - \hh_{eq} )^2 \rangle = k_BT / \chi_{\parallel}^{-1}.
\end{equation} 

Fig. \ref{fig:dqm2} shows the scaled mean squared fluctuations, which,
from \Eq{msf}, should become constant, independent of the 
system size in the thermodynamic limit for non-singular transitions.
From the figure we find this holds indeed already for the
limited system sizes studied here. Despite a considerable
scatter of the simulation data, maxima of the fluctuations for
a fixed system size are observed roughly as estimated from visual
inspection of the distributions. As an additional consistency check,
we also show as a black line the expectation for the scaled
mean squared fluctuations as obtained from the inverse
susceptibilities of the interface potential depicted in the
inset of Fig. 2(b) of the main manuscript and this document.
The agreement seems fairly good, considering the limited 
statistics and finite size effects.

In order to further analyse the rounded transitions, it is convenient to
consider the Binder cumulant of the distributions, which is given as
\cite{landau00}:
\begin{equation}
 Q_h = 1 - \frac{1}{3} \frac{ \langle ( \hh - \hh_{eq} )^4 \rangle }
 { \langle ( \hh - \hh_{eq} )^2 \rangle^2 }.
\end{equation} 

For a system away from a critical point, the distributions are Gaussian,
and $Q_h$ is approximately zero. Close to a phase transition, however,
$Q_h$ is expected to approach $2/3$ in the thermodynamic limit.

Fig. \ref{fig:cumulants} shows plots of $Q_h$ as obtained from the moments of
the global order parameter of the largest system. Results for the
remaining order parameters are similar, but are quite noisy and
do not afford a study of system size effects. From the plot,
it can be clearly seen that most values of $Q_h$ fluctuate about 0,
the result expected for a disordered phase. The location of rounded
transitions is visible in the maxima of the cumulants, as indicated
by the arrows.  The maxima are roughly coincident
with the expected location as revealed from the analysis of the
mean squared fluctuations in Fig.\ref{fig:dqm2} and the inflection points
of density profiles. However, their value is far smaller than 2/3,
the expected result for a critical point, and lend further support to our claim of
rounded layering transitions.

\section*{Density profiles for solid-like molecules}

\begin{figure*}[htb!]

   \subfigure[]{\label{perfil_sol_basal}
   \includegraphics[width=0.3\textwidth,height=0.35\paperwidth,keepaspectratio]{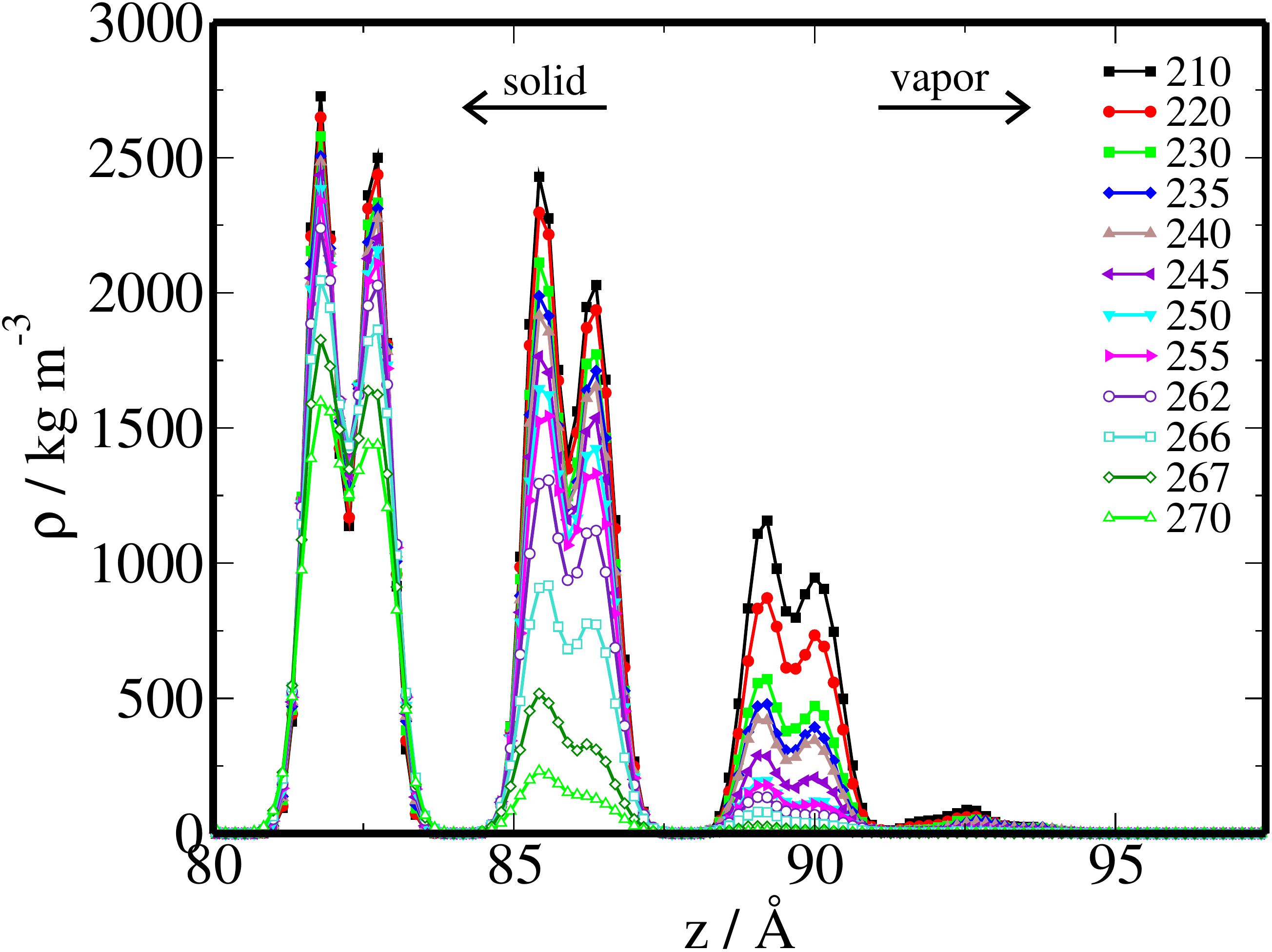}}
   \subfigure[]{\label{perfil_sol_pI}
   \includegraphics[width=0.3\textwidth,height=0.35\paperwidth,keepaspectratio]{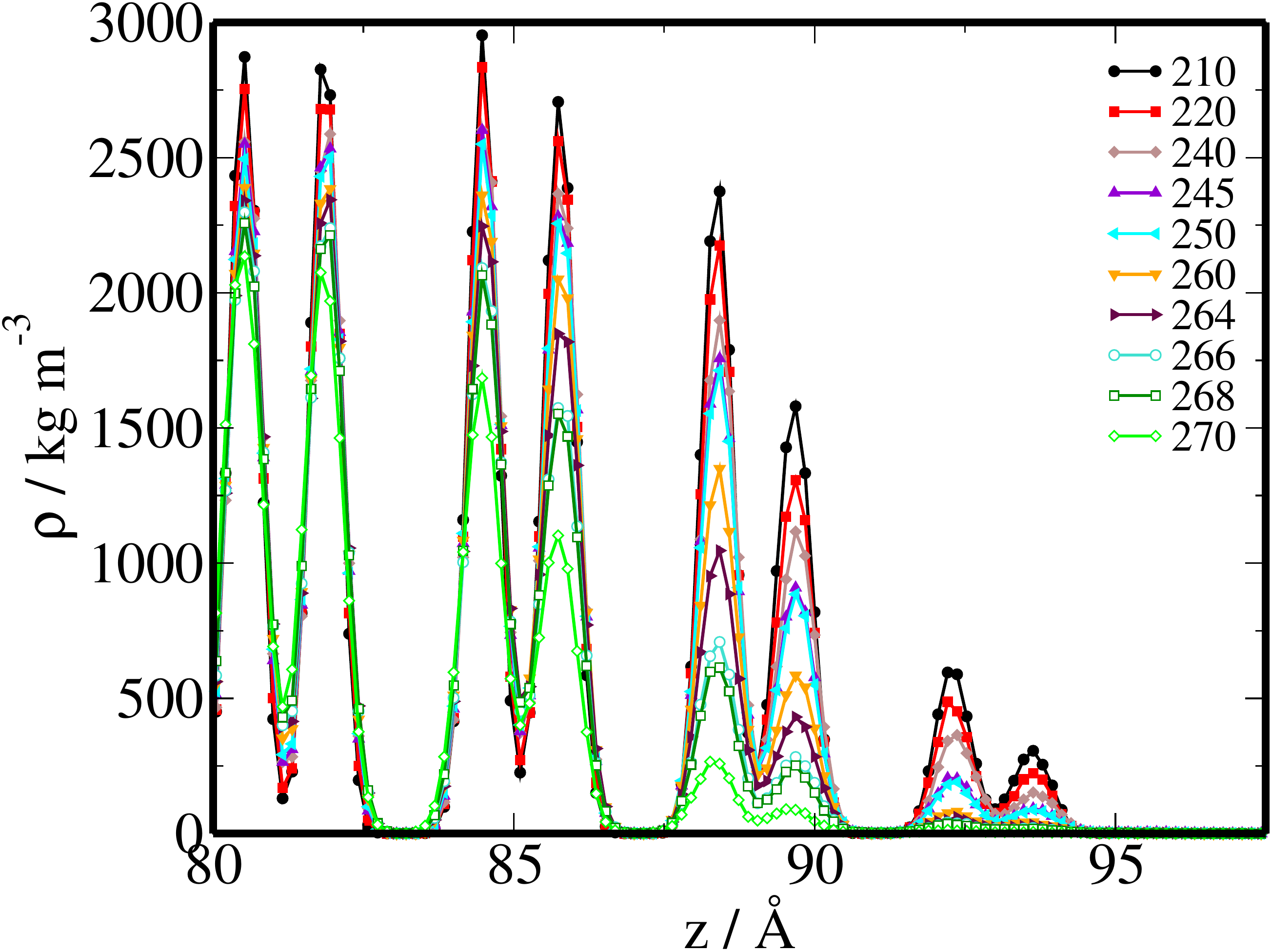}}
   \caption{Density profiles for solid-like molecules as a function of
	temperature. (a) Basal (b) Prism planes.
    }
         \label{perfiles_sup}
 \end{figure*}

Density profiles for solid-like molecules for a selected range of temperatures are shown in Fig.~\ref{perfiles_sup}.

\section*{Parameters for the fit to the interface potential}

The coefficients of the fit for the interfacial potential form given in Eq.~(\ref{eq:grn}) in the main text are displayed in Table~\ref{tab:parameters}.

\begin{table}[htb!]
 \footnotesize
  \begin{tabular}{c c c c c c c c}
  \hline
  \hline
  face & fit & A$_1$/$J\,m^{-2}$ &  A$_2$/$J\,m^{-2}$ & $\kappa \cdot$\AA & $\kappa_R$
  $\cdot$\AA &
  $k_{z,R}\cdot$\AA  &  $\theta$/rad \\
  \hline
  Basal & mf &  0.1064  &  0.00100  &  0.6170  & $2\kappa$ & 2.269  &  10.919  \\ 
  Basal & R &  0.1070   &  0.00161  &  0.6102  &  0.429  & 2.360  &  10.323  \\ 
  \hline
  pI & mf    &  0.0662 &  0.000148  &  0.6078 & $2\kappa$  &  3.149  &  7.716  \\
  pI & R    &  0.0701   &  0.00379  &  0.6206  & 0.900   &  3.144  &  7.796  \\
  \hline
  \hline
 \end{tabular}
 \caption{Parameters for the fit of simulated disjoining pressures to the model
 interface potential in  \Eq{grn} for basal and prism
 planes. Results for the constrained fit with $2\kappa=\kappa_R$ 
are also shown.}
 \label{tab:parameters}
\end{table}

\section*{  }

\section*{Computer Simulations}

   At low saturation, ice crystals grow as hexagonal prisms, exhibiting
two well defined facets. The base of the prism corresponds to
the $\{ 0 0 0 1 \}$ crystal facets, and is also known as the basal facet.
The  sides of the prism correspond to $\{ 1 0 \overline{1} 0 \}$  facets,
and are known as the prism facets. For further details on
the preparation of the initial configuration, we refer to our previous
work, Ref.\cite{benet16,benet19,llombart19}

Our simulations are carried out with the TIP4P/Ice model \cite{abascal05}.
Phase space sampling is performed using Molecular Dynamics  with
the GROMACS package. Trajectories are evolved using the
Leap-frog algorithm, with a time step of 3~fs.
Bond and angle constraints are applied using
the LINCS algorithm. Trajectories are
thermostated in the NVT ensemble using the velocity rescale 
algorithm \cite{bussi07}.
Lennard-Jones interactions are truncated at a distance of 9~{\AA}.
Electrostatic interactions are evaluated using Particle Mesh Ewald, with the
real space contribution truncated also at 9~{\AA}. The reciprocal space term 
is evaluated over a total of $80\times 64\times 160$  vectors in the
$x$, $y$, $z$ reciprocal directions, respectively. The charge structure 
factors were evaluated with a grid spacing of 0.1~nm
and  a fourth order interpolation scheme. 
For each temperature, we performed an $NpT$ simulation at 1 bar to obtain
the corresponding equilibrium lattice parameters. The $NVT$ simulations
that are then employed use equilibrated lattice parameters at that 
temperature. In order to prepare roughly square surfaces, we build
a unit supercell of size $(2\times a)\times b\times c$, with $a$, $b$ and $c$,
the unit cell parameters of a pseudo-orthorhombic cell of 8 molecules.
Simulations are performed for two system sizes, a
large system size, consisting of $8\times8\times5$ and a smaller
one with $4\times4\times5$ unit supercell with 16 molecules each. 
More detailed information on the simulations may be found
in  our previous work \cite{benet16,benet19,llombart19}.

Details of the system sizes studied may be found in 
Tables \ref{tab:system_size_basal} and \ref{tab:system_size_pI}. 
Results for the film thickness may be found in Tables \ref{tab:h_temp_basal}
and  \ref{tab:h_temp_pi}. Further details of the simulations may be found in
\cite{llombart19}.

\section*{Order parameter}

In our study, we need to distinguish solid-like from liquid-like
molecular environments in order to determine the premelting layer
thickness and to locate the ice/water surface.  
For each water molecule, we perform a neighbor analysis and
determine the $\bar q_6$ parameter, which is known to
distinguish well solid and liquid molecules in bulk
environments \cite{lechner08}. All water molecules with
$\bar q_6$ larger than a threshold value,  $\bar q_6^*$, are classified
as solid-like, while molecules with smaller $\bar q_6$ parameter
are classified as liquid-like. In order to calculate the threshold value,
we perform simulations of the bulk solid and liquid phases.
From the distribution of $\bar q_6$ in each phase, we determine
the value of $\bar q_6^*$ which produces the least amount
of mislabeling. The threshold value
is $\bar q_6^*\approx 0.35$ in the neighborhood of the triple
point, but depends slightly on temperature.
We refer to our recent work for further details and
explicit results for the dependence
of $\bar q_6^*$ on temperature \cite{llombart19}.

\begin{figure}[htb!]
   \includegraphics[width=0.40\paperwidth,keepaspectratio]{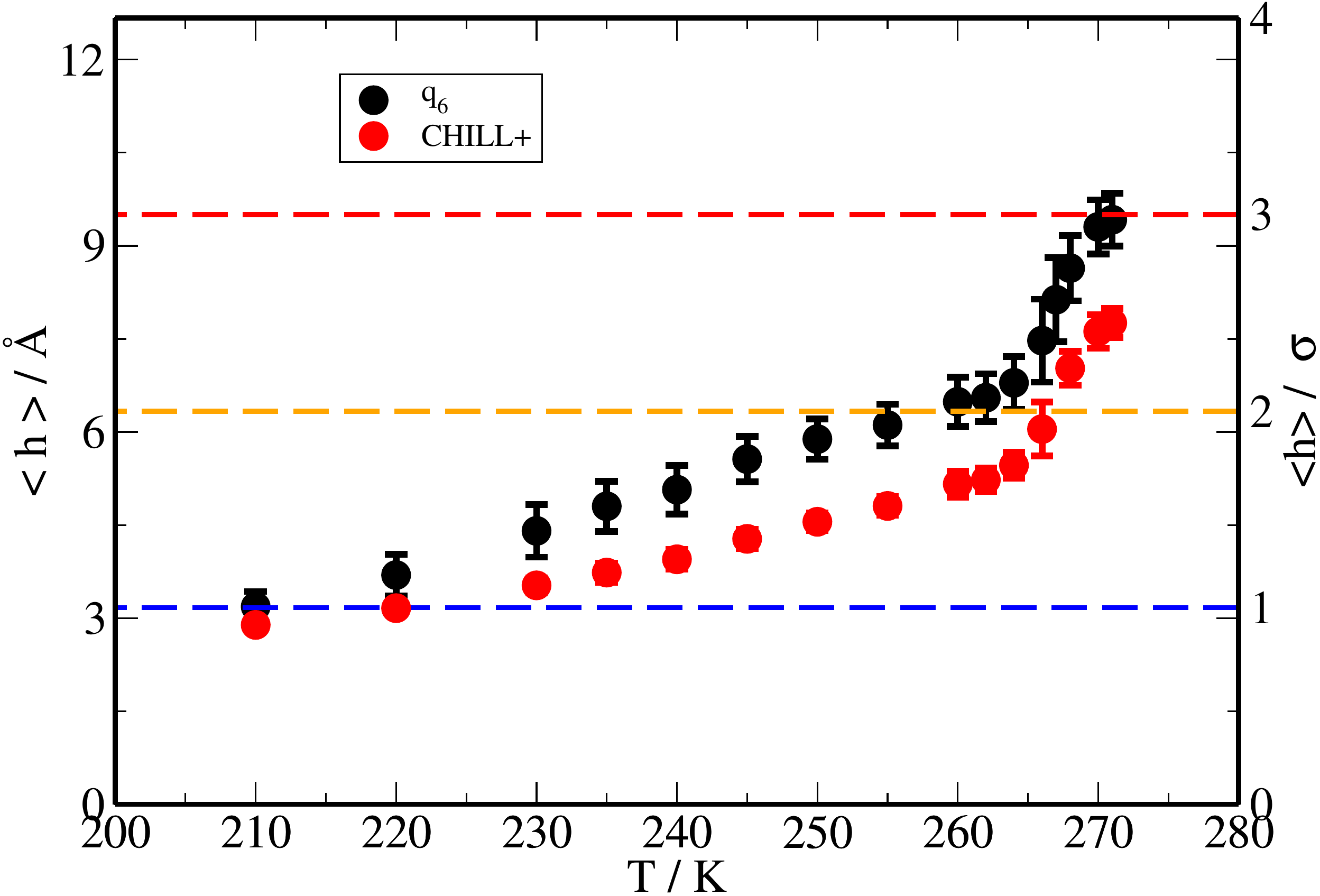}
 	     \includegraphics[width=0.40\paperwidth,keepaspectratio]{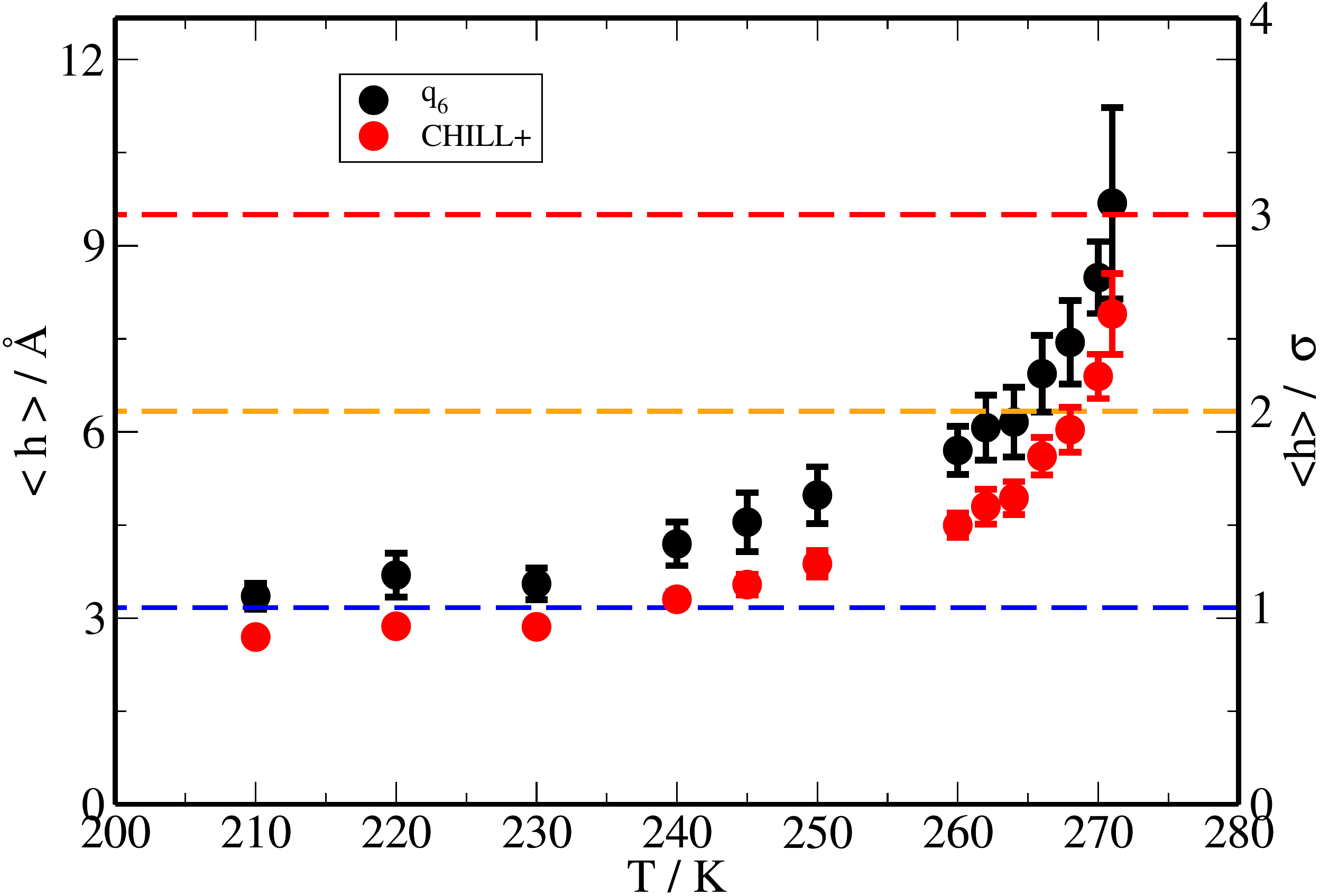}
	     \caption{Premelting layer thickness measured using the
		  $\bar q_6$ parameter as in this work (black), and with the
		  CHILL+ parameter (red) for basal (left) and prism (right) facets.
	  }
	  \label{fig:chill}
\end{figure}

An interesting alternative to the $\bar q_6$ parameter
is the CHILL+ algorithm, which has been devised explicitly
to discriminate solid-like environments different from
hexagonal ice Ih \cite{nguyen15}. In particular, the CHILL+ algorithm
allows one to identify other ice crystal allotropes such
as ice Ic, clathrate ice and interfacial ice Ih. 
In our previous work \cite{llombart19}, we have shown
that the $\bar q_6$ parameter essentially lumps all
of these forms into the disordered liquid-like category.
Therefore, our choice of order parameter essentially
distinguishes the bulk ice Ih template from all other
ice environments. The conventional application of
CHILL+ is to label solid and liquid-like molecules
in the the opposite extreme, i.e. to have
the intermediate forms ice Ic, chlathrate ice and
interfacial ice Ih lumped into the solid-like category,
and have the liquid-like category for all remaining
disordered forms. Fig. \ref{fig:chill} display
premelting film thicknesses as dictated by either
of these two prescriptions. Starting from a temperature
of about $T=240$~K, we find that either deffinitions
differ essentially by a constant offset. Below this
temperature, on the other hand, we find that both
parameters are very similar. It follows that the first
transition found in our work corresponds to an increase
of undercoordinated ice forms. Surprisingly,
both the $\bar q_6$ and  CHILL+ 
recipes yield a close to monolayer thick disordered
liquid layer at temperatures as low as 210~K.
A detailed analysis of surface ice structure and coordination 
is left for future work.

\section*{Intrinsic surfaces}

\begin{figure}[htb!]
 	     \includegraphics[width=0.50\paperwidth,height=0.50\paperwidth,keepaspectratio]{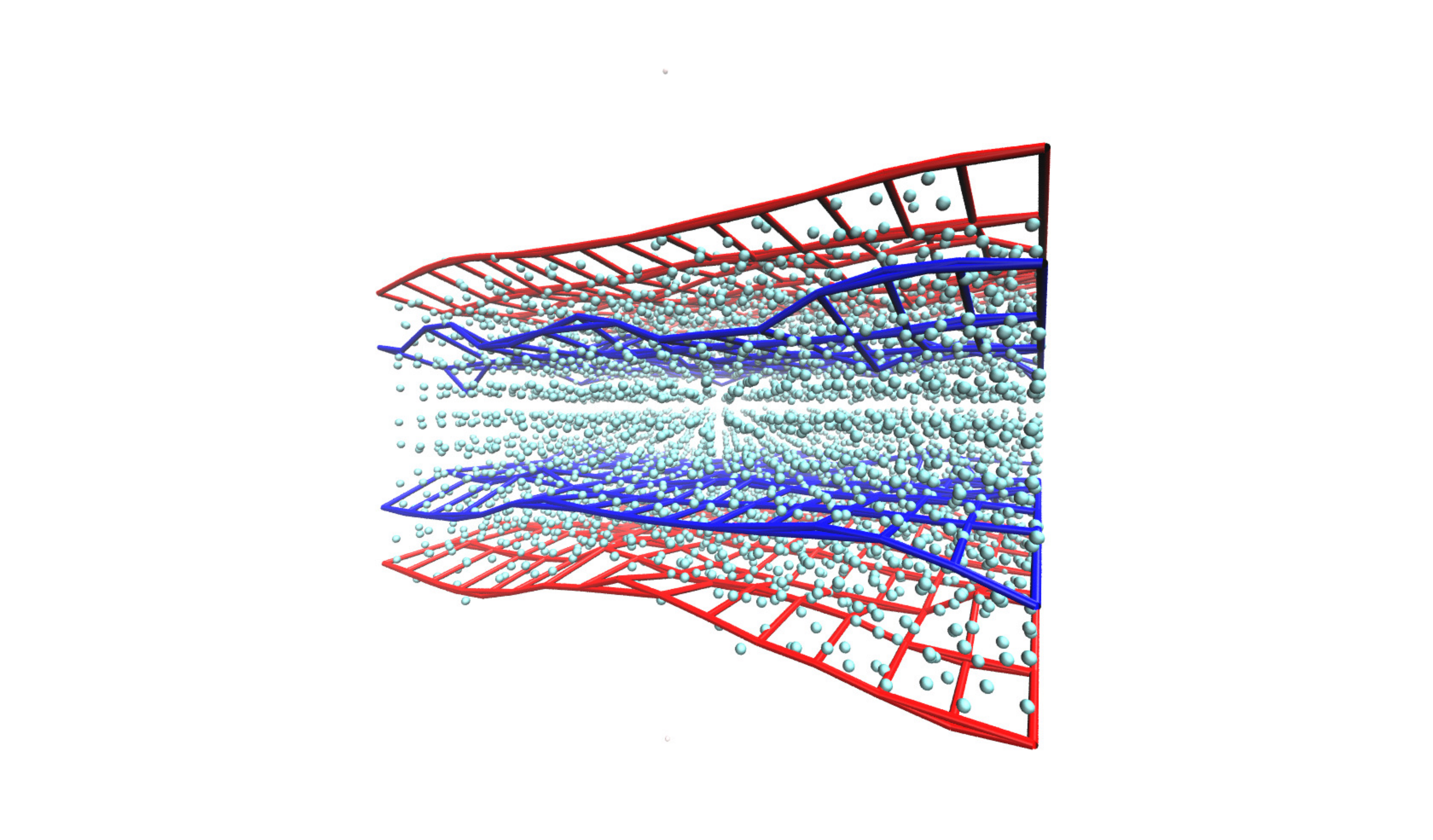}
	     \caption{Sketch showing the \SqL~and \qLV~bounding surfaces that we
	     employ to characterize the premelting film. The \SqL~surface is
	     shown in blue and the \qLV~surface is shown in red.
	  }
	  \label{mallado}
\end{figure}

For each saved snapshot, we calculate the \SqL~and \qLV~intrinsic surfaces,
meant to describe the local height of the premelting film at each point
$\rpar$ on a reference plane parallel to the interface. First we
analyze the local environment of each molecule in order to label
them as liquid-like or solid-like with the $\bar q_6$ order
parameter  as described above \cite{lechner08}. 
For a given point, $\rpar$, we calculate
the \SqL~surface as the vertical position, $\zsq(\rpar)$ obtained from
the average $z$ position of the four outermost solid-like atoms within
one pseudo-orthorhombic unit cell of size $a\times b\times c$ about the point $\rpar$. Likewise, the \qLV~surface
is calculated from the average $z$ position of the four outermost liquid-like
molecules within an area of $3\sigma\times 3\sigma$ about $\rpar$,
with $\sigma$ the Lennard-Jones range parameter of the TIP4P/Ice model. 
The procedure is illustrated in Fig.~\ref{mallado}.

\clearpage

 \section*{System sizes studied}

 Detailed information on the system sizes studied are described in Tables III
 and IV.
       
\begin{table}[h!]
 \footnotesize
  \begin{tabular}{c c c}
  \hline
  \hline
  Temperature / K & $N$ = 5120, $L_x \times L_y \times L_z$ / nm & $N$ = 1280, $L_x \times L_y \times L_z$ / nm \\
  \hline
    210  & 7.24541 $\times$ 6.27494 $\times$ 15.00000 & 3.62270 $\times$ 3.13747 $\times$ 15.00000\\
    220  & 7.24868 $\times$ 6.27776 $\times$ 15.00000 & 3.62434 $\times$ 3.13888 $\times$ 15.00000\\
    230  & 7.25247 $\times$ 6.28104 $\times$ 15.00000 & 3.62623 $\times$ 3.14052 $\times$ 15.00000\\
    235  & 7.25427 $\times$ 6.28292 $\times$ 15.00000 & 3.62714 $\times$ 3.14146 $\times$ 15.00000\\
    240  & 7.25609 $\times$ 6.28412 $\times$ 15.00000 & 3.62804 $\times$ 3.14206 $\times$ 15.00000\\
    245  & 7.25780 $\times$ 6.28587 $\times$ 15.00000 & 3.62890 $\times$ 3.14294 $\times$ 15.00000\\
    250  & 7.25949 $\times$ 6.28713 $\times$ 15.00000 & 3.62974 $\times$ 3.14356 $\times$ 15.00000\\
    255  & 7.26152 $\times$ 6.28888 $\times$ 15.00000 & 3.63076 $\times$ 3.14444 $\times$ 15.00000\\
    260  & 7.26350 $\times$ 6.29060 $\times$ 15.00000 & 3.63175 $\times$ 3.14530 $\times$ 15.00000\\
    262  & 7.26419 $\times$ 6.29120 $\times$ 15.00000 & 3.63209 $\times$ 3.14560 $\times$ 15.00000\\
    264  & 7.26488 $\times$ 6.29181 $\times$ 15.00000 & 3.63244 $\times$ 3.14590 $\times$ 15.00000\\
    266  & 7.26558 $\times$ 6.29241 $\times$ 15.00000 & 3.63270 $\times$ 3.14620 $\times$ 15.00000\\
    267  & 7.26586 $\times$ 6.29264 $\times$ 15.00000 & 3.63293 $\times$ 3.14632 $\times$ 15.00000\\
    268  & 7.26628 $\times$ 6.29302 $\times$ 15.00000 & 3.63314 $\times$ 3.14651 $\times$ 15.00000\\
    270  & 7.26698 $\times$ 6.29362 $\times$ 15.00000 & 3.63349 $\times$ 3.14681 $\times$ 15.00000\\
    271  & 7.26731 $\times$ 6.29432 $\times$ 15.00000 & 3.63360 $\times$ 3.14716 $\times$ 15.00000 \\
  \hline
  \hline
 \end{tabular}
 \caption{Temperature and systems dimensions for the basal interface. The number of unit cells in $x$, $y$ and $z$ directions are
  $N_{x}=8$, $N_{y}=8$ and $N_{z}=5$ for the bigger system. The small system has half $(N/2)$ unit cells in the $x$ and $y$ directions
  and equal number of unit cells along $z$.}
 \label{tab:system_size_basal}
\end{table}

\begin{table}[h!]
 \footnotesize
  \begin{tabular}{c c c}
  \hline
  \hline
  Temperature / K & $N$ = 5120, $L_x \times L_y \times L_z$ / nm & $N$ = 1280, $L_x \times L_y \times L_z$ / nm \\
  \hline
    210  & 7.24528 $\times$ 5.89679 $\times$ 15.00000 & 3.62270 $\times$ 2.94840 $\times$ 15.00000\\
    220  & 7.24876 $\times$ 5.89961 $\times$ 15.00000 & 3.62434 $\times$ 2.94980 $\times$ 15.00000\\
    230  & 7.25229 $\times$ 5.90249 $\times$ 15.00000 & 3.62623 $\times$ 2.95124 $\times$ 15.00000\\
    240  & 7.25604 $\times$ 5.90554 $\times$ 15.00000 & 3.62804 $\times$ 2.95277 $\times$ 15.00000\\
    245  & 7.25780 $\times$ 5.90697 $\times$ 15.00000 & 3.62890 $\times$ 2.95348 $\times$ 15.00000\\
    250  & 7.25957 $\times$ 5.90841 $\times$ 15.00000 & 3.62974 $\times$ 2.95420 $\times$ 15.00000\\
    260  & 7.26329 $\times$ 5.91143 $\times$ 15.00000 & 3.63175 $\times$ 2.95571 $\times$ 15.00000\\
    262  & 7.26405 $\times$ 5.91205 $\times$ 15.00000 & 3.63209 $\times$ 2.95602 $\times$ 15.00000\\
    264  & 7.26480 $\times$ 5.91267 $\times$ 15.00000 & 3.63244 $\times$ 2.95633 $\times$ 15.00000\\
    266  & 7.26556 $\times$ 5.91328 $\times$ 15.00000 & 3.63270 $\times$ 2.95664 $\times$ 15.00000\\
    268  & 7.26631 $\times$ 5.91389 $\times$ 15.00000 & 3.63314 $\times$ 2.95694 $\times$ 15.00000\\
    270  & 7.26707 $\times$ 5.91452 $\times$ 15.00000 & 3.63349 $\times$ 2.95726 $\times$ 15.00000\\
    271  & 7.26729 $\times$ 5.91470 $\times$ 15.00000 & 3.63360 $\times$ 2.95735 $\times$ 15.00000 \\
  \hline
  \hline
 \end{tabular}
 \caption{Temperature and systems dimensions for the prism face. The
 number of unit cells is as for the basal face listed in the caption of
 Table~\ref{tab:system_size_basal}}
 \label{tab:system_size_pI}
\end{table}

\section*{Film thickness for two different system sizes}

The film thicknesses displayed in Figures in the text are shown in tabulated
form in Tables V and VI.

\begin{table}[ht]
 \footnotesize
  \begin{tabular}{c c c}
  \hline
  \hline
  Temperature / K & $\langle h \rangle $ / \AA (Area = 1) &  $\langle h \rangle $ / \AA (Area = $\frac{1}{4}$)    \\
  \hline
    210  &  3.18 & 3.29 \\
    220  &  3.69 & 4.23 \\
    230  &  4.40 & 3.46 \\
    235  &  4.80 & 5.10 \\
    240  &  5.07 & 5.17 \\
    245  &  5.56 & 5.66 \\
    250  &  5.88 & 5.90 \\
    255  &  6.11 & 6.12 \\
    260  &  6.48 & 6.83 \\
    262  &  6.55 & 6.69 \\
    264  &  6.79 & 7.80 \\
    266  &  7.47 & 8.18 \\
    267  &  8.13 & 8.63 \\
    268  &  8.63 & 9.18 \\
    270  &  9.30 & 9.32 \\
    271  &  9.42 & 9.19 \\
  \hline
  \hline
 \end{tabular}
 \caption{ Thickness as a function of temperature for the basal face and the two systems simulated.}
 \label{tab:h_temp_basal}
\end{table}

\begin{table}[ht]
 \footnotesize
  \begin{tabular}{c c c}
  \hline
  \hline
  Temperature / K & $\langle h \rangle $ / \AA (Area = 1) &  $\langle h \rangle $ / \AA (Area = $\frac{1}{4}$)    \\
  \hline
    210  &  3.35 & 3.08 \\
    220  &  3.69 & 3.15 \\
    230  &  3.55 & 3.60 \\
    240  &  4.20 & 4.26 \\
    245  &  4.55 & 4.87 \\
    250  &  4.98 & 4.94 \\
    260  &  5.70 & 5.73 \\
    262  &  6.07 & 5.70 \\
    264  &  6.16 & 5.88 \\
    266  &  6.94 & 6.26 \\
    268  &  7.44 & 7.44 \\
    270  &  8.49 & 7.55 \\
    271  &  9.69 & 8.34 \\
  \hline
  \hline
 \end{tabular}
 \caption{ Thickness as a function of temperature for the prism face and the two systems simulated.}
 \label{tab:h_temp_pi}
\end{table}

\clearpage

\end{document}